\documentclass{article}
\usepackage{arxiv}
\usepackage[utf8]{inputenc}
\usepackage{graphicx}
\usepackage{graphics}
\usepackage{epstopdf}
\usepackage[T1]{fontenc}    
\usepackage{url}            
\usepackage{booktabs}       
\usepackage{amsfonts}       
\usepackage{nicefrac}       
\usepackage{microtype}      

\title{Detecting direct causality in multivariate time series: A comparative study}

\author{Angeliki Papana  \\
 Department of Economics\\
 University of Macedonia\\
 Egnatias 156, Thessaloniki, Greece\\
 \texttt{angeliki.papana@gmail.com, apapana@uom.edu.gr} \\
 \And
 Elsa Siggiridou \\
Department of Electrical and Computer Engineering \\ 
 Aristotle University of Thessaloniki, Greece\\
 \texttt{esingiri@ece.auth.gr} \\
 \And
 Dimitris Kugiumtzis\\
 Department of Electrical and Computer Engineering \\ 
 Aristotle University of Thessaloniki, Greece\\
 \texttt{dkugiu@auth.gr} \\
}

\begin{document}

\maketitle

\begin{abstract}
The concept of Granger causality is increasingly being applied for the
characterization of directional interactions in different applications. A
multivariate framework for estimating Granger causality is
essential in order to account for all the available information
from multivariate time series. However, the inclusion of
non-informative or non-significant variables creates
estimation problems related to the 'curse of dimensionality'. To
deal with this issue, direct causality measures using variable
selection and dimension reduction techniques have been
introduced. In this comparative work, the performance of an ensemble
of bivariate and multivariate causality measures in the time domain is assessed, 
focusing on dimension reduction causality measures.
In particular, different types of high-dimensional coupled discrete systems are used 
(involving up to 100 variables) and the robustness of the causality measures to time 
series length and different noise types is examined. The results of the simulation study 
highlight the superiority of the dimension reduction measures, especially for high-dimensional 
systems.
\end{abstract}

\keywords{Granger causality \and multivariate time
series \and variable selection \and dimension reduction \and
Monte Carlo simulations}

\section{Introduction}
Understanding the direction of information flow in coupled
systems is fundamental. The inference of causal interactions
constitutes a major issue in many real-world problems, such as
understanding the functionality of the brain
\cite{Friston11,Seth15}, inferring about the causal
relationships among different markets and market indices
\cite{Papana17,Stavroglou17} and the seismic activity
\cite{Zheng07,Chorozoglou18}.

Methods implementing the concept of Granger causality \cite{Granger69}
have been widely applied for assessing the direction of causal
effects in time series analysis. In principle, the causal effects are
inferred utilizing linear regression and prediction when the following two
criteria are satisfied: a) the
cause precedes its effect in time, and b) the causal series
contains distinct information about the series being caused that
is not available otherwise. The causality measures, as simply
termed, have implemented and extended the Granger causality in time domain
\cite{Hiemstra94,Schreiber00,Diks06,Marinazzo08,Barnett14,Papana16},
frequency domain \cite{Baccala01,Korzeniewska03,Barnett14} and
phase domain \cite{Nolte08,Nolte10,Vinck11}.

Initially, the Granger causality has been implemented
in the frame of bivariate analysis, i.e.
estimating the causal relationship in an observed variable pair
$(X,Y)$. In multivariate time series, the bivariate causality
measures may estimate indirect causality from $X$ to $Y$ stemming
from the intermediate interaction with another variable $Z$, e.g.
from the direct causal effect $X \rightarrow Z$ and $Z \rightarrow
Y$, the indirect causal effect $X \rightarrow Y$
arises. Thus, bivariate analysis cannot
distinguish between direct and indirect causal effects and may
give erroneous results when applied to multivariate systems
\cite{Blinowska04}. Therefore, it is essential to account for the
presence of the other observed variables of the multivariate time
series when testing for directional relationships between two
variables. Multivariate causality measures
utilize all the available information and aim to indicate only
direct causality \cite{Geweke82}.

Although multivariate analysis incorporates dimensional
contributions of high orders, unlike bivariate analysis, the
'curse of dimensionality' may arise, meaning that the estimation
of multivariate measures may be problematic due to the high
dimension of the observed data, leading to unreliable estimation
of direct causality \cite{Angelini10}. Over-fitting and redundancy
bias has been faced by incorporating dimension reduction
techniques \cite{Marinazzo12}. Such techniques have been developed
within the framework of subset regression
\cite{Breiman95,Collins04,Yang16}, model reduction
\cite{Bruggemann03,Shojaie10,Siggiridou16} and non-uniform
embedding \cite{Vlachos10,Faes11,Kugiumtzis13}. Variable selection
methods have also been utilized in order to limit the
dimensionality within the estimation procedure of multivariate
causality measures \cite{Marinazzo12,Songhorzadeh16}. Both
dimension reduction and variable selection methods aim to a
partially conditioning scheme.

There are various comparative studies aiming to evaluate and
compare the different causality measures. Each study focuses on
different types of measures or data characteristics. Many novel
causality measures are compared with the standard Granger
causality measures \cite{Kaminski01,Nolte10} or transfer entropy
\cite{Schreiber00}, which is the nonlinear analogue of Granger
causality, as in \cite{Kugiumtzis13,Siggiridou19}. Various works
give emphasis to the frequency-based measures
\cite{Wu11,Florin11,Fasoula13}, others concentrate on multivariate
causality measures from the time domain \cite{Papana13}. Many
causality measures have been developed in conjunction with brain
studies and there is a good number of comparison studies of
causality measures on electroencephalograms
\cite{Wu11,Florin11,Blinowska11,Silfverhuth12,Fasoula13,Olejarczyk17}.

Although different causality measures have been developed so far
and part of them aim to face the curse of dimensionality, they are
frequently tested on relatively low-dimensional simulated and real
data (i.e. systems consisted of few variables)
\cite{Guo08,Vakorin09,Runge18}, or may be applied to real data of
much higher dimensionality than the simulated ones
\cite{Zhang18,Jia19,Li20,Marinazzo12}. The influence of noise is
another factor that has not been addressed in many studies
\cite{Papana13,Jia19,Siggiridou19}. Therefore, an extensive
comparison of different causality measures regarding
high-dimensional and noisy samples is of high relevance.

In this paper, a comprehensive comparative study is performed,
including bivariate and multivariate causality measures, focusing
on measures that use dimension reduction and variable selection
techniques. For the evaluation of the measures, known coupled
systems are utilized, with different characteristics, focusing on
discrete systems. The ensemble of the optimal measures is tested
on additional simulated systems underlining the effectiveness of
the measures in high-dimensional systems (comprised up to 100
variables) and also for varying time series length
and noise types. For the assessment of the causality measures,
appropriate performance indices are employed.

The structure of the paper is as follows. In
Sec.~\ref{sec:Methodology}, we briefly present the causality
measures and significance tests, the accuracy indices for the true and estimated
causality networks and the simulation systems used in the simulation study.
In Sec.~\ref{sec:Results}, the results of the simulation study are presented and
discussed. Finally, Sec.~\ref{sec:Conclusions} summarizes the conclusions of the
study.

\section{Methodology}
\label{sec:Methodology}

In this section, the evaluation procedure of the causality
measures is discussed. First, we briefly present the ensemble of
the causality measures evaluated in this study and the
performance indices that statistically evaluate
the causality measure accuracy. Then, the synthetic systems for
the simulation analysis are described.

\subsection{Causality measures}

The examined causality measures can be grouped on the basis of whether they
are linear or nonlinear, bivariate or multivariate and for the latter whether
dimension reduction or variable selection is employed
(see Table~\ref{tab:CausalityMeasures}).
\begin{table}[ht]
\caption{List of causality measures.}
\label{tab:CausalityMeasures}
\centering
\begin{tabular}{l l l} \hline
\bf{Measures} & \bf{Linear} & \bf{Nonlinear} \\ \hline
\bf{Bivariate} & GCI \cite{Granger69} & TE \cite{Schreiber00} \\
                      &                             & MIME \cite{Vlachos10} \\ \cline{2-3} 
\bf{Multivariate} & CGCI \cite{Geweke82} & PTE \cite{Papana12} \\
& PGC \cite{Guo08} & PTERV \cite{Kugiumtzis13b} \\ \cline{2-3} 
\bf{Dimension reduction} & PCGC \cite{Marinazzo12} & PMIME \cite{Kugiumtzis13} \\
& RCGCI \cite{Siggiridou16} & PTENUE \cite{Montalto14} \\
& LFACDA \cite{Kovrenek20} & LATE \cite{Zhang18} \\
& & NLFACDA \cite{Kovrenek20} \\ \hline
\end{tabular}
\end{table}
We only consider measures defined in the time domain, here, under the assumption of
stationarity (only one of the included measures, PTERV, does not require the stationarity
assumption).
The linear causality measures are model-based measures relying
on autoregressive models, whereas the nonlinear measures are information-theoretic measures
that do not make any assumptions regarding the nature of the data. Also,
causality measures from graph theory that seek for causal parents using linear and nonlinear
correlations are considered.

Let us consider $K$ observed stationary time series. We denoted
$\{x_t\}$ the time series of the driving variable $X$, $\{y_t\}$
the time series of the response variable $Y$, whereas the other
$K-2$ observed variables are collectively denoted $Z$ and the
observations are $\mathbf{z}_t=\{z_{1,t},\ldots,z_{K-2,t}\}$, for
$t=1,\ldots,n$. Bivariate measures explore the causal
effect of the variable $X$ on $Y$ without considering the
remaining variables ($X \rightarrow Y$), while multivariate
measures account for the confounding variables of the system and
therefore consider also higher order influences ($X \rightarrow
Y|Z$). Measures using variable selection techniques are defined
conditioning on a properly extracted subset of $Z$, while measures
using dimension reduction methods exploit appropriate techniques
in order to reduce the dimensionality by selecting proper lagged
terms of the observed variables. If no otherwise
stated the implementation of the measures were done by our
research team.

\subsubsection{Model-based causality measures}

The concept of Granger Causality (GC) is the formalization of
Wiener's idea \cite{Wiener56}: if the prediction of the future
values of $Y$ is significantly improved by including past values
of another variable $X$ in the autoregressive (unrestricted) model
compared to the model that includes only past values of $Y$
(restricted model), then $X$ Granger causes $Y$ \cite{Granger69}.
Assuming that only time series $\{x_t, y_t\}$ of $X$ and $Y$ are
obtained, a bivariate vector autoregressive model (VAR) of order
$P$ is fitted (unrestricted model)
\begin{equation}
y_t = \sum_{j=1}^{P} a_{1,j}x_{t-j}+
\sum_{j=1}^{P} a_{2,j}y_{t-j} + \epsilon_t,
\label{eq:bivARfit}
\end{equation}
where $a_{1,j},a_{2,j}$ are the coefficients of the model and
$\epsilon_t$ are the residuals from fitting the model with
variance $s_U^2$. The restricted model is similarly defined but
omitting the terms of lagged variables of $X$. If the variance
$s_U^2$ of the residuals of the unrestricted model is
significantly lower than the corresponding variance $s_R^2$ of the
restricted model, then $X$ Granger causes $Y$. The magnitude of
the causal effect of $X$ on $Y$ is given by the Granger Causality
Index (GCI)
\begin{equation}
\mbox{GCI}_{X \rightarrow Y} = \ln(s_R^2 / s_U^2).
\label{eq:GCI}
\end{equation}

For multivariate time series $\{x_t, y_t, z_{1,t}, \ldots
z_{K-2,t}\}$, the GCI ignores the presence of the other $K-2$
observed variables, which may give rise to indirect causality and
positive $\mbox{GCI}_{X \rightarrow Y}$. This is the
case when the contribution of the lagged terms of $X$ is
insignificant if the lagged terms of
$\mathbf{Z}=\{Z_{1},\ldots,Z_{K-2}\}$ are included in the VAR
models. Thus, GC can be extended to assign only to
direct causal effects by conditioning on the past of all the
remaining variables of a multivariate system when forming the
restricted and unrestricted model \cite{Geweke82}. The Conditional
Granger Causality Index (CGCI) is then defined similarly to GCI in
(\ref{eq:GCI}).

The measure of Partial Granger Causality (PGC) is associated with the
concept of Granger causality and partial correlation \cite{Guo08}. It has
been introduced to address the problem of exogenous inputs and latent
variables. As in the case of the CGCI, the unrestricted and restricted models are
formed but the corresponding residual covariance matrices are assumed non-diagonal
and are decomposed to account for exogenous inputs and latent variables. In essence, PGC
constitutes an extension of the CGCI when the residuals of the VAR
models are correlated; otherwise it is identical to the CGCI.

Partially Conditioned Granger causality (PCGC) is defined
similarly to CGCI, however variable selection is utilized to
reduce the dimensionality \cite{Marinazzo12}. In particular,
conditioning is performed by inclusion of the variables in
$\mathbf{Z}$ that are the most informative for the candidate
driver variable by an information gain criterion, while the
remaining variables are excluded from the model. Thus a subset of
$K_s\cdot P$ terms of the totally $K\cdot P$ terms of the
unrestricted model in CGCI is used for the unrestricted model of
PCGC, depending on the number $K_s$ of the selected confounding
variables. Codes for the estimation of PCGC can be found in
\url{https://github.com/danielemarinazzo/PartiallyConditionedGrangerCausality}.

Restricted Conditional Granger Causality Index (RCGCI) extends the
CGCI, designed for cases with many observed variables and
relatively short time series \cite{Siggiridou16}. A modified
backward-in-time selection method is used to select the most
informative lagged terms of all $K$ variables in the unrestricted
model and only a subset of the $K\cdot P$ lagged terms enter the
unrestricted VAR model.

\subsubsection{Information-theoretic causality measures}

Transfer entropy (TE) \cite{Schreiber00} is the information-theoretic
analog of the linear Granger causality \cite{Barnett09,Schindler11}.
TE does not assume any particular model for the interactions governing the system dynamics.
Let $y_{t+1}$ be the future of $Y$,
$\textbf{x}_t = [x_t,x_{t-1},\ldots,x_{t-(m-1)\tau}]$ and
$\textbf{y}_t = [y_t,y_{t-1},\ldots,y_{t-(m-1)\tau}]$
the reconstructed vectors from each variable,
where $\tau$ is the delay time and $m$ is the embedding dimension.
The TE from $X$ to $Y$ is the mutual information of future response and the current driving state
conditioned on the current response state, $I(y_{t+1};\textbf{x}_t | \textbf{y}_t)$, and 
can be defined based on entropy terms as
\begin{equation}
\mbox{TE}_{X \rightarrow Y}=
   -H(y_{t+1}|\textbf{x}_t, \textbf{y}_t) + H(y_{t+h}|\textbf{y}_t)
\label{eq:TE}
\end{equation}
where $H(X)$ is the Shannon entropy of the variable $X$.
The entropy terms of all the information measures are estimated here using the
$k$-nearest neighbors (kNN) method, proved to be the most efficient
for high dimension \cite{Kraskov04,Papana12}. Note that in the first entropy term in
(\ref{eq:TE}) the dimension is $2m+1$.

Partial Transfer Entropy (PTE) extends TE in the multivariate case
conditioning on all the remaining variables $Z$ of the system.
Thus, the respective entropy term conditioning also on the other
$K-2$ embedded variables has dimension $K\cdot m+1$.

Mutual Information on Mixed Embedding (MIME) expands TE by
incorporating a non-uniform embedding (NUE) scheme using a subset of the
$2m$ lagged components of $\textbf{x}_t$ and $\textbf{y}_t$ assuming always
$\tau=1$, where $m$ is set to the maximum lag that may have effect on $y_{t+1}$.
Based on a conditional mutual information criterion the scheme selects the suitable
lagged terms that best explain the future of the response variable and then the normalized
conditional mutual information as in TE quantifies the causality from $X$ to $Y$
\cite{Vlachos10}.

Partial Mutual Information on Mixed Embedding (PMIME) is the
multivariate extension of MIME which accounts only for the direct
causal effects \cite{Kugiumtzis13}. Similarly to MIME, the lagged
terms are selected from all $K\cdot m$ lagged variables of the $K$
variables increasing the computation time.

Partial Transfer Entropy from a non-uniform embedding scheme
(PTENUE) is similarly defined to PMIME, changing the conditional
information criterion for the inclusion of a lagged term in NUE of
MIME and PMIME to an information criterion \cite{Montalto14}. For
the estimation of PTENUE, the ITS toolbox has been utilized:
\url{http://www.lucafaes.net/its.html}.

Partial Transfer Entropy exploiting a low-dimensional
approximation method for conditional mutual information (LATE) is
another modification of MIME and PMIME. It substitutes the
conditional mutual information in MIME and PMIME with a sum of
simple terms of mutual information and conditional mutual
information \cite{Zhang18}. LATE is suboptimal to PMIME in theory
but may offer more efficient estimation when the subset of
selected lagged terms gets large with respect to the time series
length.

Partial Transfer Entropy on Rank Vectors (PTERV) is similar to PTE but instead of the embedding vectors
$\mathbf{x}_t$ and $\mathbf{y}_t$ it uses their corresponding rank vectors \cite{Kugiumtzis13b}.
The ranks are defined regardless of the amplitude level of the embedding vector components and
in this way PTERV is robust to the presence of drifts in the time series. However, the use of the
standard estimate of relative frequency for the probability distribution and subsequently the
entropy makes the causality estimation with PTERV problematic in high dimension.

\subsubsection{Causality measures from graph theory}

Graph theory studies mathematical structures used to model
pairwise relations between objects. The causal
discovery algorithms are trying to estimate the set of such nodes
for a given target node, called the causal parent set. The linear
Fast Causal Discovery Algorithm (LFACDA) searches for causal
parents based on partial correlation, while its nonlinear
extension, denoted as NLFACDA, relies on conditional mutual
information to define the causal parents \cite{Kovrenek20}.
We used the codes for the estimation of LFACDA and
NLFACDA given by the authors in \cite{Kovrenek20} (personal
communication).

\subsection{Significance tests for the causality measures}

A causality measure gives a numerical value and one has to consider
that due to bias even a relatively high value may regard a non-existing
causal effect. Thus a significance test is conducted for the null hypothesis
that the true causality measure (if all quantities in its expression were true
and not estimated) is zero. Parametric significance tests have been
developed only for the linear model-based causality measures.
For the bivariate linear measure GCI, the null hypothesis of zero GCI is equivalent
to having all coefficients of the driving lagged term in the unrestricted model
set to zero (see (\ref{eq:bivARfit})). The following test statistic is found to follow
the Fisher-Snedecor distribution under the null hypothesis
\begin{equation}
F = \frac{(SSE_R - SSE_U)/(P_U-P_R)}{SSE_U/(n-P_R)},
\label{eq:Fisher}
\end{equation}
where $SSE_R$, $SSE_U$ are the sums of squared errors for the
unrestricted and restricted model, and $P_U$ and $P_R$ are the
number of coefficients of the unrestricted and restricted model,
respectively. The significance test for the multivariate
extensions of GCI, namely CGCI, PCGC and RCGCI, is formed
similarly by properly adapting the degrees of freedom in the
expression of the Fisher-Snedecor statistic in (\ref{eq:Fisher}).
The null distribution of the model-based PGC is not in general
known. Its significance is obtained using stationary bootstraps
(instead of surrogates), as suggested in \cite{Guo08}. For the
estimation of PGC and their statistical significance, we used the
Causal Connectivity Analysis toolbox \cite{Seth10}.

In the lack of asymptotic null distribution of the test statistic
for the nonlinear causality measures TE, PTE and PTERV, resampling
tests are developed. We use here the time-shifted surrogates and
compute the measure on an ensemble of them to form the empirical
null distribution \cite{Quiroga02}. In particular, we generate $M$
surrogate time series for the driving variable $X$, by cyclically
shifting the observations of $X$, i.e. for a randomly selected
time index $d$ the first $d$ observations of the series are moved
at the end. In this way the underlying dynamics and the marginal
distribution of $X$ are preserved in the resampled time series
while the coupling of $X$ and $Y$ is destroyed. The causality
measures are computed on the original time series and the $M$
resampled ones and the $p$-value of the one-sided significance
test is given as $p = 1 - (r_0 - 0.326)/(M+1+0.348)$, where $r_0$
is the rank of the measure value on the original data in the
ordered list of the $M+1$ measure values computed on the original
and resampled data \cite{Yu01}.

The nonlinear measures MIME, PMIME, PTENUE, and LATE involve
resampling significance test for the termination of the scheme
selecting the lagged variables. The index quantifying the driving
causal effect is zero when no lagged variables of the driver $X$
are selected, otherwise it is positive. Thus, no additional test
is required for its significance and the positive value is simply
set to one.

Finally, the measures LFACDA and NLFACDA give as output the causal
parents and therefore causal influences are inferred based on these outcomes.
Within the estimation procedure of LFACDA, the $p$-value of the partial correlation
function is assessed by a Student test, while for NLFACDA time-shifted surrogates
are utilized for assessing the significance of the conditional mutual information.

\subsection{Accuracy indexes}

For a multivariate system in $K$ variables, there are $K\cdot (K -
1)$ ordered pairs of variables to estimate causality and form the
causal network. For simulated systems, the true coupling pairs are
known, so the accuracy of any causality measure can be seen as a
binary classification accuracy, where the two classes are the
class of true coupling ordered pairs and the class of the
estimated ones converted to binary outcome (e.g. from the
significance test). To this respect, the number of true positives
(denoted as TP) and true negatives (TN) are the number of
correctly identified ordered pairs having and not having causal
relationship, respectively, whereas false negatives (FN) and false
positives (FP) correspond to ordered coupling pairs having causal
relationships that were not detected and not having causal
relationship but being detected as such, respectively.

In this study, we consider three binary classification indexes: a)
the sensitivity, given as the proportion of the true causal
effects correctly identified as such, Sens = TP/(TP + FN), b) the
specificity given as the proportion of the pairs correctly not
identified as being coupled, Spec = TN/(TN + FP), and c) the
F1-score quantifying the overall performance given as the harmonic
mean of precision and recall (sensitivity), F1 = 2 $\cdot$ precision $\cdot$
sens/(precision+sens), where precision = TP/(TP+FP) is the
proportion of correctly identified positive cases from all the
predicted positive cases. The three indexes Sens, Spec and F1 are
actually expressed in percentages, where the highest value $100\%$
regards perfect match of the pairs of causal effect, the pairs of
non-causal effect, and all pairs, respectively.

\section{Numerical results for artificial time series}
\label{sec:Results}

In this section, we report the results of the simulation analysis.
The considered multivariate systems exhibit various desired
characteristics and known connectivity patterns. Having the actual
and the estimated couplings of a system from each measure, the
accuracy indexes of sensitivity (Sen), specificity (Spec), and
F1-score (F1) are calculated.

The evaluation of the performance of the causality measures is
presented for the different simulation systems. The ranking of the
measures is performed based on the mean F1-score from all
considered systems, realizations, time series lengths and coupling
strengths. Sensitivity and specificity are displayed in order to
obtain a more comprehensive insight on the effectiveness of each
measure.

We set the number of realizations of each simulation system to be
equal to 100 for systems with $K=4$ and 5 variables, 30
realizations for $K=10$, and 10 realizations for $K>10$ due to the
computational cost. The simulated time series are stationary as
required for the majority of the causality measures. In addition
to the variety of the systems characteristic as described above,
we want to monitor the performance of measures as the data
dimension, length of time series and the strength of the couplings
increase. The considered time series lengths are
$n=512,1024,2048,4096$.

The free parameters for the computation of the causality measures
are the standard values as commonly used in
previous reported works. For TE and PTE, the time lag $\tau$ is
equal to one (as in the original definition of TE in
\cite{Schreiber00}). The embedding dimension $m$ (for TE, PTE) is
equal to the maximum delay shown in the equations of each
simulation system and the same stands for the order $P$ of the VAR
(for GCI, CGCI, PCGC, PGC). We also set the number of neighbor's
$k$ equal to 10 (for information theoretic measures). For PCGC,
the number of conditioning variables is apriori defined for each
system, after taking into consideration the plateau of information
gain by adding more variables in the estimation of the measure (as
described in \cite{Marinazzo12}). For example, for systems with
$K=4$ variables, we set $K_s=1$, for systems with $K=5$, we set
$K_s=2$. The selection of $K_s$ for the remaining systems is
discussed below. The maximum lag $L_{max}$ for the estimation of
the information measures based on the non-uniform embedding scheme
does not affect the estimation if it is sufficiently large, 
i.e.~it should be equal or larger than the true maximum delay occurring
in the equations of each simulation system. In particular, we set
$L_{max} = 5$ for all the considered systems.

\subsection{Preliminary simulation results from the ensemble of causality measures}

The ensemble of the causality measures, 14 in total, are evaluated
on artificial time series from systems with different properties.
From the ensemble of the causality measures, the optimal causality measures are then
determined based on the mean F1-score from the selected artificial systems.
In the considered simulation systems, linear and nonlinear couplings, as well as unidirectional
and bidirectional couplings exist. Both stochastic and chaotic systems are assumed.
The simulation systems are analytically presented below.

\begin{figure}[!h]
\centering
 \includegraphics[scale=.32]{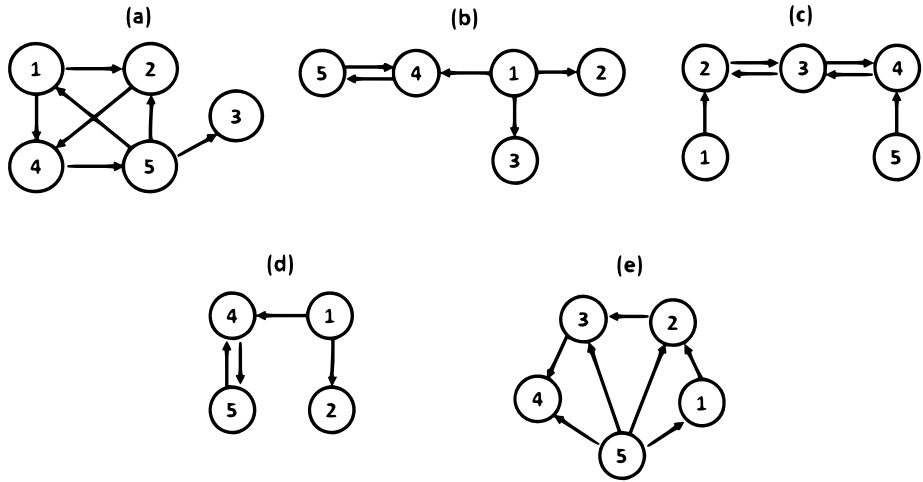}
 \caption{The true causal network of (a) S1, (b) S2, (c) S3 in $K=5$ variables, (d) S4, (e) S5. }
 \label{fig:SystNetworks}
\end{figure}

\textbf{(S1)} The first simulation system is a VAR(4) process in five variables with unidirectional
causal relationships:
$X_1 \rightarrow X_2$, $X_1 \rightarrow X_4$, $X_2 \rightarrow X_4$, $X_4 \rightarrow X_5$,
$X_5 \rightarrow X_1$, $X_5 \rightarrow X_2$ and $X_5 \rightarrow X_3$ \cite{Schelter06}
\begin{eqnarray}
  x_{1,t} & = & 0.4x_{1,t-1} - 0.5x_{1,t-2} + 0.4x_{5,t-1} + \epsilon_{1,t} \nonumber\\
  x_{2,t} & = & 0.4x_{2,t-1} - 0.3x_{1,t-4} + 0.4x_{5,t-2} + \epsilon_{2,t} \nonumber\\
  x_{3,t} & = & 0.5x_{3,t-1} - 0.7x_{3,t-2} - 0.3x_{5,t-3} + \epsilon_{3,t} \nonumber\\
  x_{4,t} & = & 0.8x_{4,t-3} + 0.4x_{1,t-2} + 0.3x_{2,t-3} + \epsilon_{4,t} \nonumber\\
  x_{5,t} & = & 0.7x_{5,t-1} - 0.5x_{5,t-2} - 0.4x_{4,t-1} + \epsilon_{5,t}  \nonumber
\label{eq:Schelter}
\end{eqnarray}
where $\epsilon_{i,t}$, $i=1,\ldots,5$ are independent to each other Gaussian white noise
processes with unit standard deviation (Fig.~\ref{fig:SystNetworks}a).

\textbf{(S2)} The second system has linear ($X_1 \rightarrow X_3$,
$X_4 \leftrightarrow X_5$) and nonlinear couplings ($X_1
\rightarrow X_2$,  $X_1 \rightarrow X_4$) as given by the
equations:
\begin{eqnarray}
  x_{1,t} & = & 0.95 \sqrt{2} x_{1,t-1} - 0.9025 x_{1,t-2} + \epsilon_{1,t} \nonumber\\
  x_{2,t} & = & 0.5 x_{1,t-2}^2 + \epsilon_{2,t}  \nonumber\\
  x_{3,t} & = & -0.4x_{1,t-3} + \epsilon_{3,t} \nonumber\\
  x_{4,t} & = & -0.5 x_{1,t-2}^2 + 0.25 \sqrt{2} x_{4,t-1} + 0.25 \sqrt{2} x_{5,t-1} + \epsilon_{4,t}, \nonumber\\
  x_{5,t} & = & -0.25 \sqrt{2} x_{4,t-1} + 0.25 \sqrt{2} x_{5,t-1} + \epsilon_{5,t}, \nonumber
\label{eq:Montalto}
\end{eqnarray}
with Gaussian noise terms as in S1 \cite{Montalto14} (Fig.~\ref{fig:SystNetworks}b).

\textbf{(S3)} Finally, we consider a nonlinear dynamical system,
the coupled H{\'e}non maps (discrete time) given by the equations
\begin{eqnarray}
 x_{i,t} & = & 1.4 - x_{i,t-1}^2 + 0.3x_{i,t-2}, \quad i=1,K  \nonumber\\
 x_{i,t} & = & 1.4 - \left(0.5c(x_{i-1,t-1}+x_{i+1,t-1})+(1-c)x_{i,t-1}\right)^2+ 0.3x_{i,t-2},  \quad  i=2,\ldots,K-1 \nonumber
 \label{eq:coupledHenon}
\end{eqnarray}
\cite{Politi92}. The system can be defined for an arbitrary number of variables $K$,
while the parameter $c$ controls the coupling strength between the
variables. For this system, different scenarios are considered.
First, we set the number of variables equal to $K=5$ and the coupling strength varies
as $c=0.1,0.2,0.3,0.4,0.5$ (Fig.~\ref{fig:SystNetworks}c). Then, we raise the dimensionality
of the system and set $K=10$ and $K=15$ while the coupling strength is fixed to $c=0.3$
(intermediate coupling strength).

\subsubsection{Results for bivariate causality measures}
By definition, bivariate measures (GCI, TE, MIME) indicate both
direct and indirect causal effect. The GCI, which is linear,
outperforms the nonlinear ones in case of the linear stochastic
system S1 but performs poorer in case of the nonlinear systems.
Since the performance of the measures is assessed on the basis of
the true network including only direct causal effects, the
bivariate measures are 'penalized' for indicating indirect effects
which is not fair. Therefore, we examined carefully the
performance of these measures by looking at the individual
extracted causal effects. The low specificity of the three
bivariate measures, which affects their overall performance, is
partly because of the detection of indirect couplings and partly
due to false indications of causal relationships.

Particularly, in S1, the bivariate measures achieve high
sensitivity, correctly finding the true causal effects, however
their specificity is low. Indirect couplings are indicated with
one or more intermediate variables, e.g.~TE indicates the indirect
causal effect $X_2 \rightarrow X_3$ with 2 intermediate variables
($X_2 \rightarrow X_4 \rightarrow X_5 \rightarrow X_3$). The GCI
indicates the spurious links $X_3 \rightarrow X_2$, $X_3
\rightarrow X_4$, $X_3 \rightarrow X_5$, and MIME falsely suggests
$X_3 \rightarrow X_2$ and $X_3 \rightarrow X_4$. Only TE does not
suggest any false causal effects.

Regarding S2, the GCI performs poorer than the nonlinear measures;
indicates the nonlinear causal effects $X_1 \rightarrow X_2$ and
$X_1 \rightarrow X_3$ with relatively low percentages over the 100
realizations, e.g.~for $n=2048$ the percentages
for these two causal effects are $19\%$ and $33\%$, respectively.
The nonlinear measures TE and MIME reveal all the true couplings.
Apart from the indirect links, the three measures suggest also
false causal influences and therefore all have a low specificity
that affects their overall performance.

In case of S3, the measures achieve again a high sensitivity since
they derive the true connections. The GCI and TE suggest the
spurious links $X_2 \rightarrow X_1$ and $X_4 \rightarrow X_5$,
however the low specificity of MIME is only due to the detection
of indirect causality. The increase of the dimensionality should
not affect the performance of the bivariate measures, however the
declining specificity as $K$ grows is due to the increase of the
indirect connections in the system. Table~\ref{tab:bivariate}
summarizes the results of the bivariate measures from the
simulation analysis.
\begin{table}[ht]
 \caption{Sensitivity, specificity and F1-score for the bivariate causality measures on
 the simulation systems S1, S2, S3 in $K=5$ variables (coupling strength $c=0.1 -0.5$),
 S3 in $K=10$ variables ($c=0.3$), and S3 in $K=15$ variables ($c=0.3$) as a mean over all time 
 series lengths.}
 \label{tab:bivariate}
 \centering
 \begin{tabular}{l lll | l lll } \hline
  \bf{GCI}         & \bf{Sens} & \bf{Spec} & \bf{F1-score} & \bf{TE}  & \bf{Sens} & \bf{Spec}  & \bf{F1-score} \\ \hline
  S1                  & 100          & 26.21       & 59.66             & S1                 & 99.61       & 59.41        & 73.13    \\
  S2                  & 68.65       & 68            & 51.79             & S2                 & 94.7         & 30.48        & 47.08    \\
  S3, $K=5$      & 92.93       & 66.71        & 71.4              & S3, $K=5$      & 99.05       & 72.54        & 76.65    \\
  S3, $K=10$    & 96.62       & 55.53        & 48.9              & S3, $K=10$    & 100          & 49.81        & 46.84    \\
  S3, $K=15$    & 97.31       & 50.96        & 36.61            & S3, $K=15$    & 100          & 43.18        & 33.76    \\ \hline
 \end{tabular}
 \begin{tabular}{l lll} \hline
  \bf{MIME}      & \bf{Sens} & \bf{Spec}  & \bf{F1-score}  \\ \cline{1-4}
  S1                  & 100          & 46.37        & 67.31    \\
  S2                 & 92.95        & 44.17        & 51.69    \\
  S3, $K=5$      & 98.02       & 77.39        & 79.44     \\
  S3, $K=10$    & 99.48       & 64             & 55.07     \\
  S3, $K=15$    & 99.52       & 66.36        & 46.05     \\ \cline{1-4}
\end{tabular}
\end{table}

\subsubsection{Results for fully conditioned causality measures}

Fully conditioned measures (CGCI, PGC, PTE, PTERV) are designed to infer only about direct
causality (and not indirect ones)
since they consider all the observed variables of the system within their estimation procedure. Therefore,
they are expected to have an increased specificity compared to the bivariate ones which are 'penalized' for indicating
indirect causal effects. As the dimensionality of the examined systems increases, the multivariate measures are expected to have a decreasing performance due to the curse of dimensionality. Table~\ref{tab:conditioned} summarizes the outcomes of the simulation study for the fully
conditioned measures.
\begin{table}[ht]
 \caption{Sensitivity, specificity and F1-score for the multivariate causality measures on
 the simulation systems S1, S2, S3 in $K=5$ variables (coupling strength $c=0.1 -0.5$),
 S3 in $K=10$ variables ($c=0.3$), and S3 in $K=15$ variables ($c=0.3$) as a mean over all time 
 series lengths.}
 \label{tab:conditioned}
 \centering
 \begin{tabular}{l lll |l lll } \hline
  \bf{CGCI}    & \bf{Sens} & \bf{Spec} &\bf{F1-score} &\bf{PGC}  & \bf{Sens} & \bf{Spec}& \bf{F1-score}  \\ \hline
  S1               & 100          & 94.81       & 95.69       & S1                & 98.86       & 99.45       & 98.89    \\
  S2               & 61.7         & 79.33       & 54.85       & S2                & 58.1         & 85.32       & 57.83    \\
  S3, $K=5$   & 93.21       & 67.44       & 70.16       & S3, $K=5$     & 70.27       & 84.46       & 64.41     \\
  S3, $K=10$ & 95            & 69.32       & 57.3        & S3, $K=10$    & 82.94       & 74.56       & 56.88     \\
  S3, $K=15$ & 95.19       & 72.09        & 49.13      & S3, $K=15$    & 69.91      & 86.29       & 53.15    \\ \hline
  \bf{PTE}  & \bf{Sens}  & \bf{Spec}   & \bf{F1-score} & \bf{PTERV}  & \bf{Sens} & \bf{Spec}   & \bf{F1-score}  \\ \hline
  S1               & 93         & 91.14         & 88.86        & S1                 & 11.82     & 79.39         & 10.36    \\
  S2               & 81.25     & 76.97         & 65.23        & S2                & 33.55      & 96.02         & 52.56    \\
  S3, $K=5$   & 84.54     & 83.8           & 73.07        & S3, $K=5$    & 84.47      & 79.33         & 71.38     \\
  S3, $K=10$ & 81.3       & 78.08         & 60.49        & S3, $K=10$   & 51.2       & 89.9           & 50.12     \\
  S3, $K=15$ & 46.83     & 96.41         & 53.88        & S3, $K=15$   & 19.9       & 81.1           & 15.95     \\ \hline
\end{tabular}
\end{table}

The linear measures CGCI and PGC are very effective in case of S1, achieving high scores and outperforming
the nonlinear PTE. The PTE requires longer time series length to detect all the direct links with high
percentages over the 100 realizations while the indirect link $X_5 \rightarrow X_4$ arises as the time series
length increases. The PTERV has a really poor performance with exceptionally low sensitivity.

In case of S2, the linear measures perform poorer than PTE since they have a difficulty in detecting
the two nonlinear causal effects of the system.
Although PGC is designed for systems with exogenous inputs and latent variables, it does not achieve a 
high score. All but the PTERV measures indicate spurious links giving relatively low specificity. The PTERV 
has instead a very low sensitivity having overall poor performance with F1 score somewhat smaller than for 
the linear measures.

For system S3, the performance of all measures worsens with the increase of the number of variables $K$. 
The PTE succeeds the highest mean F1-score among the four fully
conditioned causality measures. PTE suggests some indirect links but with relatively low percentages
while it spuriously detects the links
$X_2 \rightarrow X_1$ and $X_{K-1} \rightarrow X_K$ for all $K$.
The CGCI indicates the true connections with an increasing sensitivity with the time series length.
However, it shows more indirect and spurious causal effects than PTE.
The PGC also indicates indirect and spurious causal links while it requires longer time series
and/or strongly coupled variables to correctly infer about the links.
PTERV has its best performance in this simulation study for S3 with $K=5$, but as $K$ increases,
its performance gets poor again.

\subsubsection{Results for multivariate measures with dimension reduction}

Multivariate measures with dimension reduction have been introduced for high-dimensional systems
to address the curse of dimensionality.
Further, depending on the complexity of the scheme to reduce the dimension, the computational cost 
may be smaller than for the fully conditioned measures, where the resampling significance test is also 
included. It is noted that LATE is computationally more demanding compared to PMIME and PTENUE, 
especially in case of high dimensions and large time series lengths.  Table~\ref{tab:partiallyconditioned} 
displays summary results of these causality measures on the systems S1, S2 and S3. 
\begin{table}[ht]
 \caption{Sensitivity, specificity and F1-score for the multivariate causality measures using dimension reduction on the simulation systems S1, S2, S3 in $K=5$ variables (coupling strength $c=0.1 -0.5$),
 S3 in $K=10$ variables ($c=0.3$), and S3 in $K=15$ variables ($c=0.3$) as a mean over all time series lengths.}
 \label{tab:partiallyconditioned}
 \centering
 \begin{tabular}{l lll | l lll} \hline
 \bf{PCGC} & \bf{Sens}  & \bf{Spec}  & \bf{F1-score}  & \bf{RCGCI}     & \bf{Sens}  & \bf{Spec}  & \bf{F1-score} \\ \hline
  S1                & 100           & 92.83         & 94.12            & S1                  & 100          &  95.67       & 96.39            \\
  S2                & 61.65        & 78.65         & 54.47            & S2                  & 61.85        &  82.4        & 57.02            \\
  S3, $K=5$     & 84.64        & 71.68         & 67.73            & S3, K=5          & 87.36        & 81.17        & 75.48           \\
  S3, $K=10$   & 83.7          & 81.91         & 62.04            & S3, K=10        & 91.56        &  77.18       & 62.08            \\
  S3, $K=15$   & 82.98        & 88.23         & 59.09            & S3, K=15        & 90.19        & 79.55        & 52.93            \\ \hline
 \bf{LFACDA} & \bf{Sens}  & \bf{Spec}   & \bf{F1-score} & \bf{NLFACDA} & \bf{Sens} & \bf{Spec} & \bf{F1-score} \\ \hline
  S1                & 99.89       & 98.06          & 98.28            &  S1          & 45.57        & 92.12        & 56.04    \\
  S2                & 57.85       & 85.17          & 56.79            &  S2          & 77.05        & 92.79        & 77.36    \\
  S3, $K=5$     & 74.86       & 87.49          & 71.34            &  S3, $K=5$     & 14.98        & 98.5         & 21.55    \\
  S3, $K=10$   & 79.43       & 75.1            & 54.69            &  S3, $K=10$   & 6.8            & 99.44       & 11.83    \\
  S3, $K=15$   & 60.48       & 86.63          & 47.22            &  S3, $K=15$   &  5.7           & 99.38       & 10.08   \\  \hline
 \bf{PMIME} & \bf{Sens} & \bf{Spec}    & \bf{F1-score} & \bf{PTENUE}  & \bf{Sens}  & \bf{Spec}  & \bf{F1-score}  \\ \hline
  S1                & 99.97       & 87.67          & 90.22            &   S1               & 99.79        & 97.54        & 97.79    \\
  S2                & 93.1        & 87.05           & 80.91            &   S2               & 91.9          & 88.67        & 81.75    \\
  S3, $K=5$     & 96.65      & 99.74           & 97.47            & S3, $K=5$      & 94.98        & 99.93        & 96.2   \\
  S3, $K=10$   & 99.69      & 99.63           & 99.03            & S3, $K=10$    & 99.48        & 99.7          & 99.06   \\
  S3, $K=15$   & 96.24      & 99.57           & 96.2             & S3, $K=15$     & 99.52        & 99.58        & 98.33   \\  \hline
\end{tabular}
 \begin{tabular}{l lll } \hline
  \bf{LATE}    & \bf{Sens} & \bf{Spec}  & \bf{F1-score} \\ \cline{1-4}
  S1               & 93.61      & 90.1          & 88.49    \\
  S2               & 87.6        & 70.12        & 63.76    \\
  S3, $K=5$    & 94.79      & 90.95        & 87.77     \\
  S3, $K=10$  & 92.4        & 94.18        & 84.04     \\
  S3, $K=15$  & 90.87      & 94.16        & 78.79     \\  \cline{1-4}
\end{tabular}
\end{table}

The linear measures PCGC\footnote{Regarding the number of conditioning variables $K_S$ for the estimation of PCGC we 
set $K_S=2$ for S1, S2 and S3 in 5 variables, $K_S=3$ for S3 in 10 variables and $K_S=4$ for S3 in 15 variables.},
RCGCI and LFACDA are all very effective in case of the linear system S1.
However, as expected, a poorer performance is observed in case of the nonlinear systems.
Regarding S2, the linear measures hardly detect the two nonlinear links, while they suggest
spurious and indirect causal effects.
Their performance on S3 with $K=5$ is somewhat better than for S2 but worsens as the dimensionality of S3 increases ($K=10$ and $K=15$).
The sensitivity of RCGCI stays high, however indirect links are obtained.
For large time series lengths and large coupling strengths, the specificity and therefore
the overall efficacy of the measure is affected by the detection of spurious links.
The PCGC performs slightly worse than RCGCI, since spurious influences are
detected also for smaller time series lengths and weaker coupling strengths. LFACDA
has the smallest sensitivity among the three linear partially conditioning measures but
similar specificity.

The PTENUE, PMIME and LATE exploit the non-uniform embedding scheme in order to reduce the
dimensionality of the required estimations. PTENUE has similar performance with the linear measures
for S1, while PMIME closely follows scoring a bit lower due to the detection of non-coupled pairs of variables
with percentages a bit larger than the considered nominal level ($5\%$).
LATE has a slightly lower score due to its lower sensitivity for small time series lengths, although it achieves
a bit higher specificity than PMIME.
The NLFACDA has the poorest performance among the measures of dimension reduction due to its low
sensitivity. 

In Table~\ref{tab:meanF1Scores}, we give an overall score of the three accuracy indexes for all the causality 
measures computed as averages of the respective accuracy indexes on the simulation
systems S1, S2, S3 in $K =5$ variables ($c = 0.1,0.2,0.3,0.4,0.5$), S3 in $K =10$ variables ($c = 0.3$) and 
S3 in $K =15$ variables ($c = 0.3$) considering all the time series lengths.  
\begin{table}[ht]
 \caption{Overall score of accuracy indexes for the causality measures from
 the three first simulation systems, as a mean over all realizations and all time series lengths. 
Measures are listed in descending order with respect to the mean F1-score.}
 \label{tab:meanF1Scores}
 \centering
 \begin{tabular}{l lll} \hline
 \bf{Measure} & \bf{Sens}  & \bf{Spec}  & \bf{F1-score} \\ \hline
 PTENUE         & 97.13        &  97.08       & 94.63             \\
 PMIME           & 97.13        &  94.73       & 92.77             \\
 LATE             &  91.85       &  87.90       & 80.57             \\
 RCGCI            &  86.19       & 83.19       & 68.78             \\
 PTE               &  77.38       &  85.28       & 68.31             \\
 PCGC             &  82.59       & 82.66        & 67.49             \\
 PGC               &  76.02       &  86.02       & 66.23             \\
 LFACDA         &  74.50       &  86.49       & 65.66             \\
 CGCI              &  89.02       &  76.60       & 65.43             \\
 MIME             &  97.99       &  59.66       & 59.91             \\
 TE                 &  98.67       &  51.08       & 55.49             \\
 GCI                &  91.10       &  53.48       &  53.67             \\
 PTERV           &  40.19        &  85.14      & 40.07            \\
 NLFACDA       & 30.03         &  96.45      & 35.37            \\ \hline
\end{tabular}
\end{table}
The causality measures are displayed in descending order based on the mean extracted F1-score.
PTENUE and PMIME outperform all the other measures, achieving very high scores.
LATE closely follows PTENUE and PMIME obtaining slightly lower scores due to the suboptimal 
approximation of the conditional mutual information that takes place within the estimation
procedure of the measure. The two linear measures scoring highest are RCGCI and PCGC, 
which also apply dimension reduction.

\subsection{Additional simulation results from the optimal causality measures}

Based on the preliminary simulation study, we focus on the two optimal linear (RCGCI, PCGC) and the
two optimal nonlinear causality measures (PTENUE, PMIME) and further explore their efficiency in
detecting the causal influences of multivariate systems. Therefore, results from additional simulations
systems with different characteristics are reported. In particular, we consider a system with a missing
variable (S4), a system with a common driver (S5), a nonlinear high-dimensional stochastic system (S6),
the chaotic system (S3) where the dimensionality is gradually increased ($K$ goes from 20 to 100 with a 
step of 10). Further, we examine the influence of noise on the measures by considering observational noise 
of different types and levels.

\subsubsection{Nonlinear stochastic VAR, the case of missing variable}   

We consider the simulation system S2, whereas the variable $X_3$ is excluded from
the analysis when seeking to determine the causal effects. Therefore, S4 includes the
remaining 4 variables ($X_1,X_2,X_4,X_5$) of S2. Since $X_3$ is not influencing any other variables,
the true causal network of S2 is expected to be as for S4 but without the link $X_1 \rightarrow X_3$
(Fig.~\ref{fig:SystNetworks}d). The results are shown in Table~\ref{tab:S4results} for different time 
series lengths.
\begin{table}[ht]
 \caption{Accuracy indexes for the measures PTENUE, PMIMEm RCGCI and PCGC on the simulation system S4 and different time series lengths.}
 \label{tab:S4results}
 \centering
 \begin{tabular}{l lll | l lll} \hline
 \bf{PTENUE} & \bf{Sens}  & \bf{Spec}  & \bf{F1-score} &  \bf{PMIME} & \bf{Sens}  & \bf{Spec}   & \bf{F1-score} \\ \hline
 $n=512$        & 77.25       &  90.13       & 78.83              &  $n=512$        & 78          & 86.12         & 76.7                \\
 $n=1024$      & 83.5         & 84.88        & 67.74              & $n=1024$      & 88.75      & 85.75         & 82.16              \\
 $n=2048$      &  99.25      & 76             & 80.83              & $n=2048$      & 99.75      & 82.75         & 85.99              \\
 $n=4096$      & 100         & 73.75         & 79.51              & $n=4096$      & 100        & 79.25         & 83.52              \\
 MEAN            & 90           & 81.19         & 79.39              & MEAN            & 91.63     & 83.47         & 82.09              \\ \hline
 \bf{RCGCI} & \bf{Sens}  & \bf{Spec}   & \bf{F1-score}   & \bf{PCGC} & \bf{Sens}    & \bf{Spec}   & \bf{F1-score} \\ \hline
 $n=512$        & 43          & 73.62         & 42.36              &  $n=512$        & 46.25     & 68.5           & 42.69              \\
 $n=1024$      & 50          & 70              & 46.19              &  $n=1024$      & 55.5       & 66.37         & 49.02              \\
 $n=2048$      & 59.25     & 68.87          & 52.69              &  $n=2048$      & 63.25     & 64.12         & 53.02              \\
 $n=4096$      & 58          & 63.62          & 49.57              & $n=4096$      & 61         & 60.25          & 49.99              \\
 MEAN            & 52.56     & 69.03          & 47.7                & MEAN             & 56.5      & 64.81          & 48.68              \\ \hline
\end{tabular}
\end{table}

The nonlinear measures correctly identify the true causal links, however the linear ones do not indicate the
two nonlinear links and therefore have a much lower sensitivity.
The sensitivity of all the examined measures rises with the time series length.
PMIME and PTENUE indicate the spurious causal effects $X_2 \rightarrow X_1$ and $X_2 \rightarrow X_3$.
RCGCI and PCGC (we set $K_S=1$) falsely suggest the links $X_2 \rightarrow X_3$, $X_4 \rightarrow X_2$ 
and $X_5 \rightarrow X_2$.
The specificity of all the measures decreases with the time series length $n$ since the percentage of the erroneously
significant causal effects over the 100 realizations grows with $n$.
Overall, PMIME outperforms the other measures for system S4 achieving the highest mean F1-score and
PTENUE closely follows.
 
\subsubsection{Nonlinear chaotic system with a common driving variable}   

Next, we consider S5, a nonlinear chaotic system in 5 variables, where $X_5$ is a common driver
for the remaining variables of the system
(Eq.5 from \cite{Koutlis19})
\begin{eqnarray}
x_{1,t} & = & 1.4 - s x_{5,t-1}^2 -(1-s) x_{1,t-1}^2 + 0.3x_{1,t-2}   \nonumber\\
x_{i,t} & = & 1.4 - s x_{5,t-1}^2 - c x_{i-1,t-1} x_{i,t-1} - (1-c-s) x_{i,t-1}^2 + 0.3 x_{i,t-2}, i=2,3,4 \nonumber \\
x_{5,t} & = & 1.4 - x_{5,t-1}^2 + 0.3x_{5,t-2}. \nonumber
 \label{eq:toymodelK5}
\end{eqnarray}
The coupling strength among the variables is $c=0.1$ and the driving strength of the source variable $X_5$ is set to $s=0.1$ (Fig.~\ref{fig:SystNetworks}e).

The chaotic nature of S5 favors the nonlinear measures, as shown in Table~\ref{tab:S5results}. 
\begin{table}[ht]
 \caption{Accuracy indexes for the measures PTENUE, PMIMEm RCGCI and PCGC on the simulation system S5 and different time series lengths.}
 \label{tab:S5results}
 \centering
 \begin{tabular}{l lll |l lll } \hline
 \bf{PTENUE} & \bf{Sens} & \bf{Spec}  & \bf{F1-score} & \bf{PMIME} & \bf{Sens}  & \bf{Spec}  & \bf{F1-score} \\ \hline
 $n=512$        & 66           &  100          & 79.14             &  $n=512$     & 77          & 99.54         & 86.17               \\
 $n=1024$      & 93           &  100          & 96.17             &  $n=1024$      & 97          & 99.85         & 98.24            \\
 $n=2048$      & 100         &  100          & 100                &  $n=2048$      & 99.75     & 100            & 100               \\
 $n=4096$      & 100         &  100          & 100                &  $n=4096$      & 100        & 100            & 100               \\
 MEAN            & 89.75      &  100          & 93.83             &  MEAN            & 93.5       & 99.85         & 96.1               \\ \hline
 \bf{RCGCI} & \bf{Sens}  & \bf{Spec}   & \bf{F1-score} & \bf{PCGC} & \bf{Sens}    & \bf{Spec}  & \bf{F1-score}  \\ \hline
 $n=512$        & 56.43     & 92.85         & 66.21             &  $n=512$        & 56.71     & 95.23         & 68.09              \\
 $n=1024$      & 65.86     & 92.77         & 73.44             &  $n=1024$      & 66.71     & 93.62         & 74.68              \\
 $n=2048$      & 71.29     & 91.54         & 76.32             &  $n=2048$      & 79.43     & 91.62         & 81.48              \\
 $n=4096$      & 83.14     & 88              & 81.2              &   $n=4096$      & 93.86     & 86.31         & 86.11              \\
 MEAN            & 69.18     & 91.29         & 74.29            & MEAN             & 74.18     & 91.7           & 77.59              \\ \hline
\end{tabular}
\end{table}
Indeed, PTENUE and PMIME outperform the linear measures scoring very high and do not seem to be affected 
by the common source in case of S5. It is clear that sufficient large data size is required so that the measures 
achieve their optimal performance. 
The efficacy of RCGCI and PCGC (we set $K_S=2$) is improved with the time series length as a result of the 
increase of the sensitivity of the measures. On the other hand, their specificity decreases with the data size 
since spurious links increase as well, such as $X_1 \rightarrow X_5$ and $X_4 \rightarrow X_5$.

\subsubsection{Nonlinear VAR in $K=20$ variables, with variable square and product terms}   

The simulation system S6 is a multivariate nonlinear VAR model of order 3 in $K=20$
variables, defined as
\begin{eqnarray}
x_{1,t} & = & 0.8 x_{1,t-1} - 0.1 x_{3,t-2} - 0.3 x_{19,t-2} + e_1t   \nonumber\\
x_{2,t} & = & 0.8 x_{2,t-1} - 0.2 x_{1,t-2} x_{15,t-1} + e_{2,t}  \nonumber \\
x_{3,t} & = & 0.8 x_{3,t-1}  + e_{3,t}  \nonumber \\
x_{4,t} & = & 0.8 x_{4,t-1} - 0.2 x_{5,t-2} x_{7,t-1} + e_{4,t}  \nonumber \\
x_{5,t} & = & 0.8 x_{5,t-1} - 0.5 x_{3,t-1}^2  + e_{5,t}  \nonumber \\
x_{6,t} & = & 0.8 x_{6,t-1} - 0.4 x_{3,t-3} + 0.1 x_{10,t-1}^2 + e_{6,t}  \nonumber \\
x_{7,t} & = & 0.8 x_{7,t-1} + e_{7,t}  \nonumber \\
x_{8,t} & = & -0.6 x_{8,t-1} + 0.1 x_{11,t-3} x_{19,t-2} + e_{8,t}  \nonumber \\
x_{9,t} & = & 0.8 x_{9,t-1} - 0.1 x_{20,t-1}  + e_{9,t}  \nonumber \\
x_{10,t} & = & 0.8 x_{10,t-1}  + e_{10,t}  \nonumber \\
x_{11,t} & = & 0.8 x_{11,t-1} + 0.1 x_{20,t-1} -  0.1 x_{4,t-2} ^2 + e_{11,t}  \nonumber \\
x_{12,t} & = & 0.8 x_{12,t-1} + 0.2 x_{19,t-2} - 0.1 x_{16,t-1}^2 + e_{12,t}  \nonumber \\
x_{13,t} & = & 0.8 x_{13,t-1} + e_{13,t}  \nonumber \\
x_{14,t} & = & 0.8 x_{14,t-1} - 0.2 x_{4,t-3}^2 + e_{14,t}  \nonumber \\
x_{15,t} & = & 0.8 x_{15,t-1}  + e_{15,t}  \nonumber \\
x_{16,t} & = & 0.8 x_{16,t-1} + 0.1 x_{19,t-2} - 0.3 x_{3,t-2} + e_{16,t}  \nonumber \\
x_{17,t} & = & 0.8 x_{17,t-1} - 0.1 x_{7,t-2}^2 + e_{17,t}  \nonumber \\
x_{18,t} & = & 0.8 x_{18,t-1} - 0.3 x_{7,t-2} + 0.5 x_{2,t-3} x_{20,t-1} + e_{18,t}  \nonumber \\
x_{19,t} & = & 0.8 x_{19,t-1}  + e_{19,t}  \nonumber \\
x_{20,t} & = & 0.8 x_{20,t-1}  + e_{20,t}  \nonumber
\label{eq:nLinearK20}
\end{eqnarray}
where $e_{i,t},i=1,\ldots,20$ are independent normal white noise processes with unit standard deviation.
The system equations contain terms of squares and products of the variables, which does not favor the linear 
measures RCGCI and PCGC that by construction cannot handle such relationships. The connectivity network of 
S6 contains 23 true links and 357 uncoupled pairs of variables.

The linear measures detect much fewer true causal influences than the nonlinear ones, since they do not manage 
to infer about all nonlinear causal effects, such as $X_1 \rightarrow X_2$, $X_3 \rightarrow X_5$ and 
$X_5 \rightarrow X_4$. Therefore a rather low sensitivity is accomplished, as shown in Table~\ref{tab:S6results}.
\begin{table}[ht]
 \caption{Accuracy indexes for the measures PTENUE, PMIMEm RCGCI and PCGC on the simulation system S6 and different time series lengths.}
 \label{tab:S6results}
 \centering
 \begin{tabular}{l lll | l lll} \hline
 \bf{PTENUE} & \bf{Sens} & \bf{Spec}  & \bf{F1-score} & \bf{PMIME} & \bf{Sens}   & \bf{Spec} & \bf{F1-score} \\ \hline
 $n=512$        & 72.61       &  93.73       & 55.11             &  $n=512$        & 82.17      & 80.25         & 33.68           \\
 $n=1024$      & 84.78       &  93.64       & 61                  &  $n=1024$      & 87.83      & 82.75         & 39.24           \\
 $n=2048$      & 89.13       &  91.74       & 59.08             &  $n=2048$      & 93.91      & 83.67         & 42.68           \\
 $n=4096$      & 91.3         & 92.72        & 62.56             &  $n=4096$      & 95.22      & 85.94         & 46.61           \\
 MEAN            & 84.46       & 92.96        & 59.44             &  MEAN            & 89.78      & 83.15         & 40.55           \\ \hline
 \bf{RCGCI} & \bf{Sens}  & \bf{Spec}   & \bf{F1-score} &  \bf{PCGC} & \bf{Sens}    & \bf{Spec}  & \bf{F1-score} \\ \hline
 $n=512$        & 51.74     & 94.48         & 43.55             &  $n=512$        & 45.22     & 92.52         & 34.84             \\
 $n=1024$      & 61.3       & 96.02         & 54.82             &  $n=1024$      & 62.17     & 93.28         & 46.63             \\
 $n=2048$      & 55.22     & 96.02         & 50.78             &  $n=2048$      & 58.7       & 92.66         & 43.02             \\
 $n=4096$      & 58.7       & 96.05         & 53.4              &  $n=4096$      & 65.65     & 91.93         & 45.09              \\
 MEAN            & 56.74     & 95.64         & 50.64            & MEAN             & 57.94     & 92.6           & 42.4               \\ \hline
\end{tabular}
\end{table}
Further, RCGCI and PCGC (we set $K_S=3$) indicate much fewer false positives compared to the
nonlinear measures. On the other hand, PTENUE and PMIME have a rather high sensitivity that increases 
with the time series length.

Although nonlinear links are correctly identified, some linear links (such as
$X_7 \rightarrow X_{18}$ and $X_{20} \rightarrow X_{11}$) are not always found.
PMIME suggests the most spurious and indirect links, and a few less are found by PTENUE. Indicatively,
the mean TP, TN, FP and FN over the 10 realizations of S6 with $n=2048$ are displayed in
Table~\ref{tab:S6links}.
\begin{table}[ht]
 \caption{Mean estimated TP, TN, FP, FN over the 10 realizations of simulation system S6
with $n=2048$.}
 \label{tab:S6links}
 \centering
 \begin{tabular}{l llll} \hline
 \bf{Measure} & \bf{TP}  & \bf{TN}  & \bf{FP}   & \bf{FN} \\ \hline
 \bf{PTENUE} & 20.5       & 327.5      & 29.5       & 2.5         \\
 \bf{PMIME}   & 21.6       & 298.7      & 58.3       & 1.4         \\
 \bf{RCGCI}   & 12.7       & 342.8       & 14.2       & 10.3        \\
 \bf{PCGC}     & 13.5       & 330.8      & 26.2       & 9.5          \\ \hline
\end{tabular}
\end{table}

\subsubsection{Chaotic system, increasing dimensionality}    
The simulation system S3 is suitable for observing the performance of the measures in case of chaotic
systems and also for evaluating the measures while increasing the dimensionality. Therefore, we vary
the number of variables of S3 from $K=20$ up to $K=100$ with a step of 10. The coupling strength is
fixed to $c=0.3$.

\begin{figure}[ht]
\centering
\includegraphics[scale=.29]{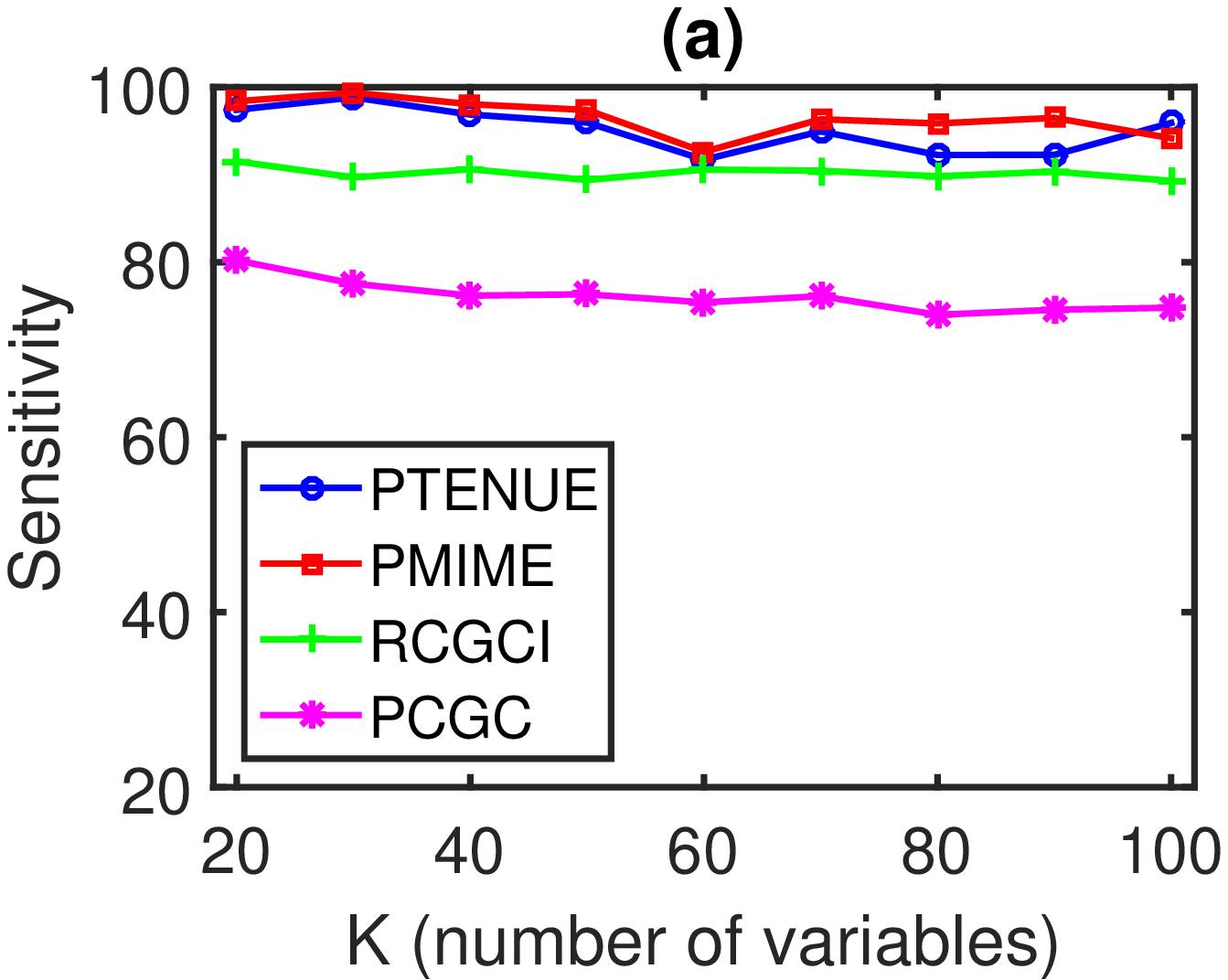}
\includegraphics[scale=.29]{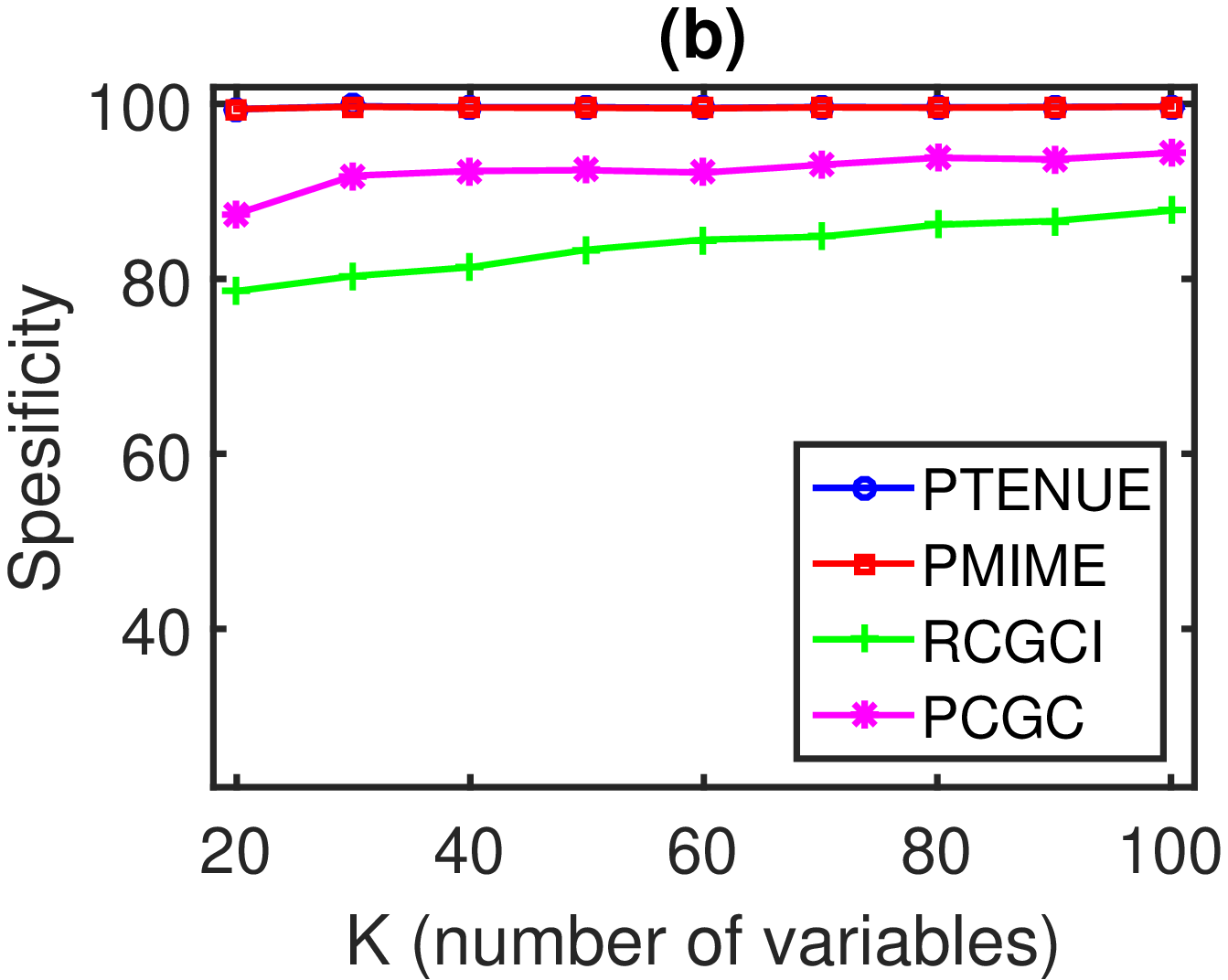}
\includegraphics[scale=.29]{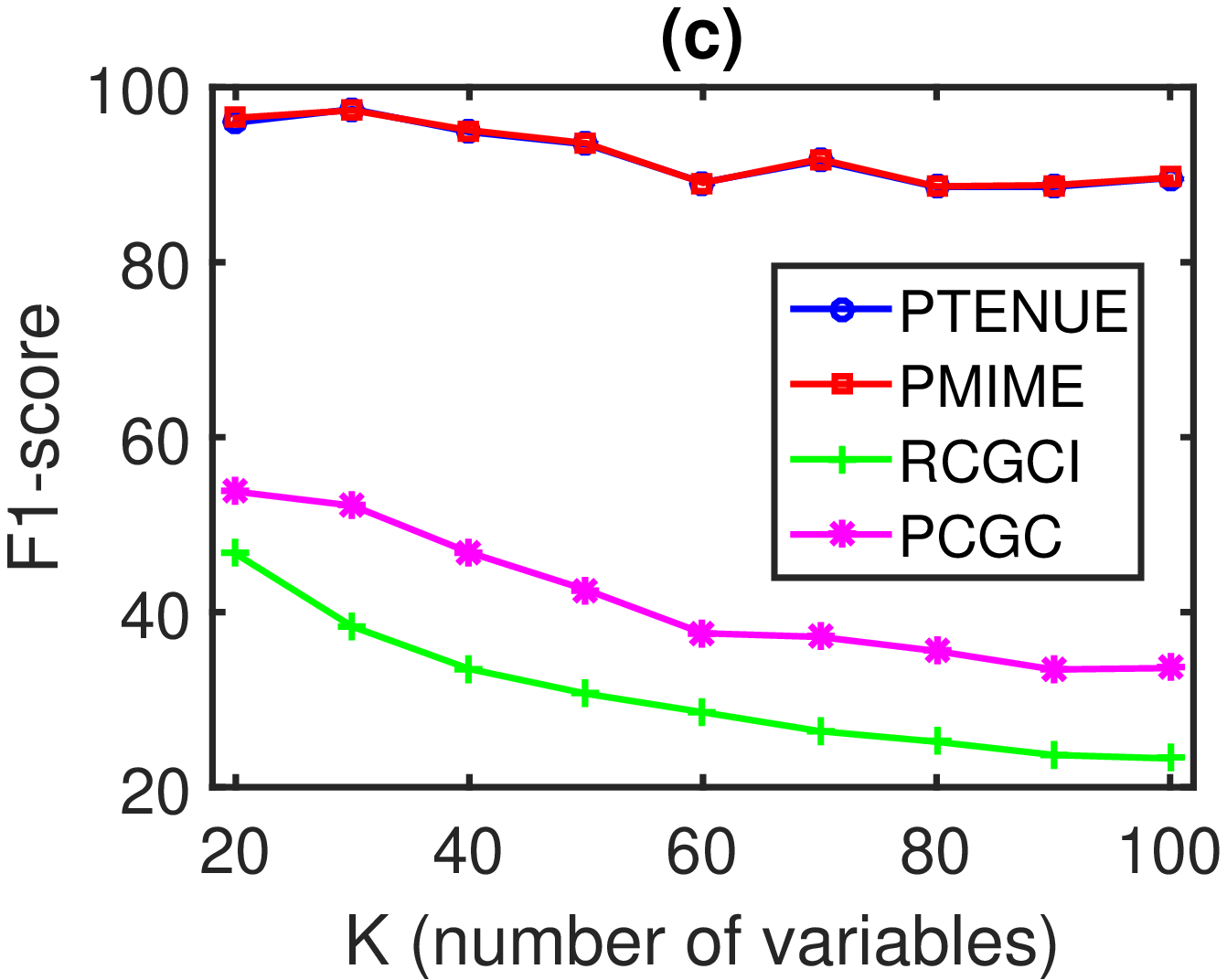}
 \caption{Accuracy indexes for S3 as a function of the number of variables $K$ as a mean over
all time series lengths.}
 \label{fig:S3increasingK}
\end{figure}

\begin{figure}[!h]
\centering
 \includegraphics[scale=.25]{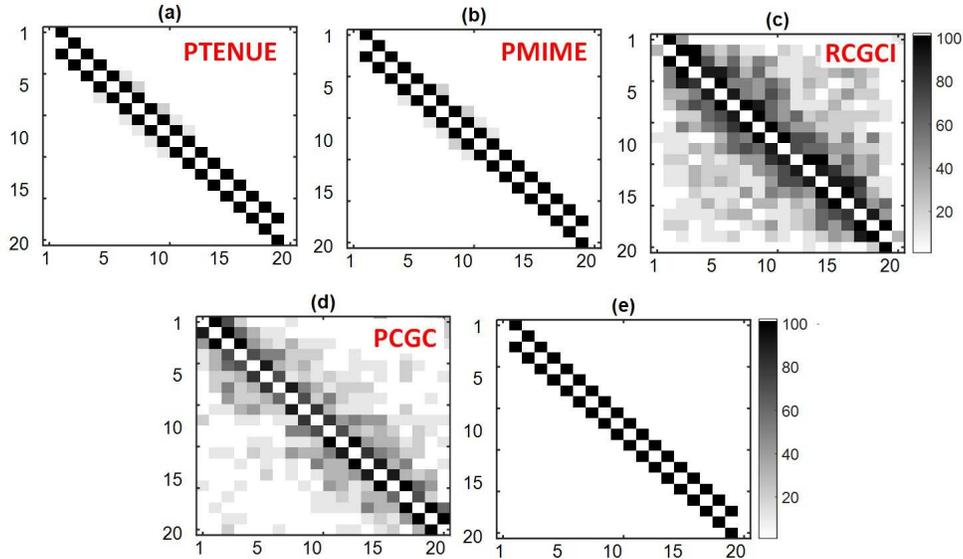}
 \caption{The heat matrix (gray scale) of the causality measure for all variable pairs of system S3 in $K=20$ 
 variables with $n=2048$: (a) PTENUE, (b) PMIME, (c) RCGCI, (d) PCGC, (e) the true causal network.}
 \label{fig:S3K20causality}
\end{figure}

PTENUE and PMIME correctly detect the causal links, even for
the high dimensional cases. Few indirect links with low percentages may appear. Their sensitivity increases
with the time series length while specificity stays at the same level. This is shown in Table~\ref{tab:S3K20binary} 
for $K=20$. 
\begin{table}[ht]
 \caption{Accuracy indexes for the causality measures PTENUE, PMIME, RCGCI, PCGC from the simulation system S3 in $K=20$ variables with $c=0.3$.}
 \label{tab:S3K20binary}
 \centering
 \begin{tabular}{l lll | l lll} \hline
 \bf{PTENUE} & \bf{Sens} & \bf{Spec}  & \bf{F1-score} & \bf{PMIME} & \bf{Sens}   & \bf{Spec} & \bf{F1-score} \\ \hline
 $n=512$        & 92.22       & 98.78       & 90.48             &  $n=512$        & 95           & 98.72       & 91.81           \\
 $n=1024$      & 98.06       & 99.3         & 95.9               &  $n=1024$      & 99.17      & 99.39        & 96.85           \\
 $n=2048$      & 100          & 99.65       & 98.42             &  $n=2048$      & 100         & 99.65        & 98.41           \\
 $n=4096$      & 99.44       & 99.85       & 99.04             &  $n=4096$      & 99.44      & 99.83        & 98.92           \\
 MEAN            & 97.43       & 99.4         & 95.96             &  MEAN            & 98.4        & 99.4          & 96.5           \\ \hline
 \bf{RCGCI} & \bf{Sens}  & \bf{Spec}   & \bf{F1-score} &  \bf{PCGC} & \bf{Sens}    & \bf{Spec}  & \bf{F1-score} \\ \hline
 $n=512$        & 84.44     & 82.21         & 48.03             &  $n=512$        & 65.83      & 89.3          & 49.88             \\
 $n=1024$      & 88.61     & 81.37         & 48.77             &  $n=1024$      & 74.72      & 88.52         & 52.97             \\
 $n=2048$      & 94.72     & 76.63         & 45.57             &  $n=2048$      & 85.83      & 87.3           & 56.56             \\
 $n=4096$      & 98.33     & 74.19         & 44.29             &  $n=4096$      & 94.44      & 84.59         & 55.72              \\
 MEAN            & 91.53     & 78.6           & 46.67             & MEAN             & 80.21      & 87.43         & 53.78             \\ \hline
\end{tabular}
\end{table}
The performance of the nonlinear measures does not significantly change as the dimensionality of S3
increases, as shown in Fig.~\ref{fig:S3increasingK}.
PMIME has the highest sensitivity, which stays at high levels for all $K$.
PTENUE closely follows achieving also a high sensitivity for all $K$.
The specificity of PTENUE is slightly highest than that of PMIME. A high specificity is achieved for the nonlinear measures
that seems to be unaffected by the increase of the dimensionality.
Overall, a high mean F1-score for all $K$ reflects the effectiveness of the two nonlinear measures.
Indicatively, the extracted causal effects for S3 in $K=20$ variables with $n=2048$ based on the
four measures is shown in Fig.~\ref{fig:S3K20causality}, whereas also the percentage of detection
of each causal link over all the realizations of the system is displayed.

The linear measures have a much inferior performance compared to the nonlinear ones.
The influence of the time series length is more pronounced for the linear measures; their sensitivity increases 
with the time series length while specificity is reduced (see Table~\ref{tab:S3K20binary}).
RCGCI and PCGC (we increase $K_S$ from 4 ($K=20$) to 18 ($K=100$)) indicate the true causal influences, however 
suggest various indirect links as well as the spurious links $X_2 \rightarrow X_1$ and $X_{K-1} \rightarrow X_K$.
Their sensitivity is at a lower level than that of the nonlinear measures since they detect the causal links at a lower 
percentage over all the realizations. PCGC obtains the lowest sensitivity.
Their specificity is also poorer compared to PMIME's and PTENUE's, but slightly increases with $K$ due to the 
decreasing percentages of indirect causal effects (see Fig.~\ref{fig:S3increasingK}).
Overall, a very low mean F1-score is reached by the linear measures which decreases with $K$.

\subsubsection{Chaotic system with observational noise}    
The effect of noise on the four causality measures is examined by adding observational noise
to the simulation system S3. In particular, we consider S3 in $K=5$ variables, fix the coupling strength to $c=0.2$ and change the noise levels to $10\%, 20\%$ and $50\%$ (the noise standard deviation is a percentage of the noise-free data standard deviation). Moreover, different types of noise are considered: Gaussian white noise, white noise from t-Student distribution with 2 degrees of freedom (heavy tailed distribution) and noise given by a GARCH(1,1) system exhibiting volatility clusterings. The latter is defined as
\begin{eqnarray}
  e_t & = & \sigma_t w_t    \nonumber\\
  \sigma_t^2 & = & 0.2 + 0.2 e_{t-1}^2 + 0.75 \sigma_{t-1}^2 \nonumber
 \label{eq:GARCH11}
\end{eqnarray}
where $w_t$ is a Gaussian white noise process.

\begin{figure}[!h]
\centering
\includegraphics[scale=.26]{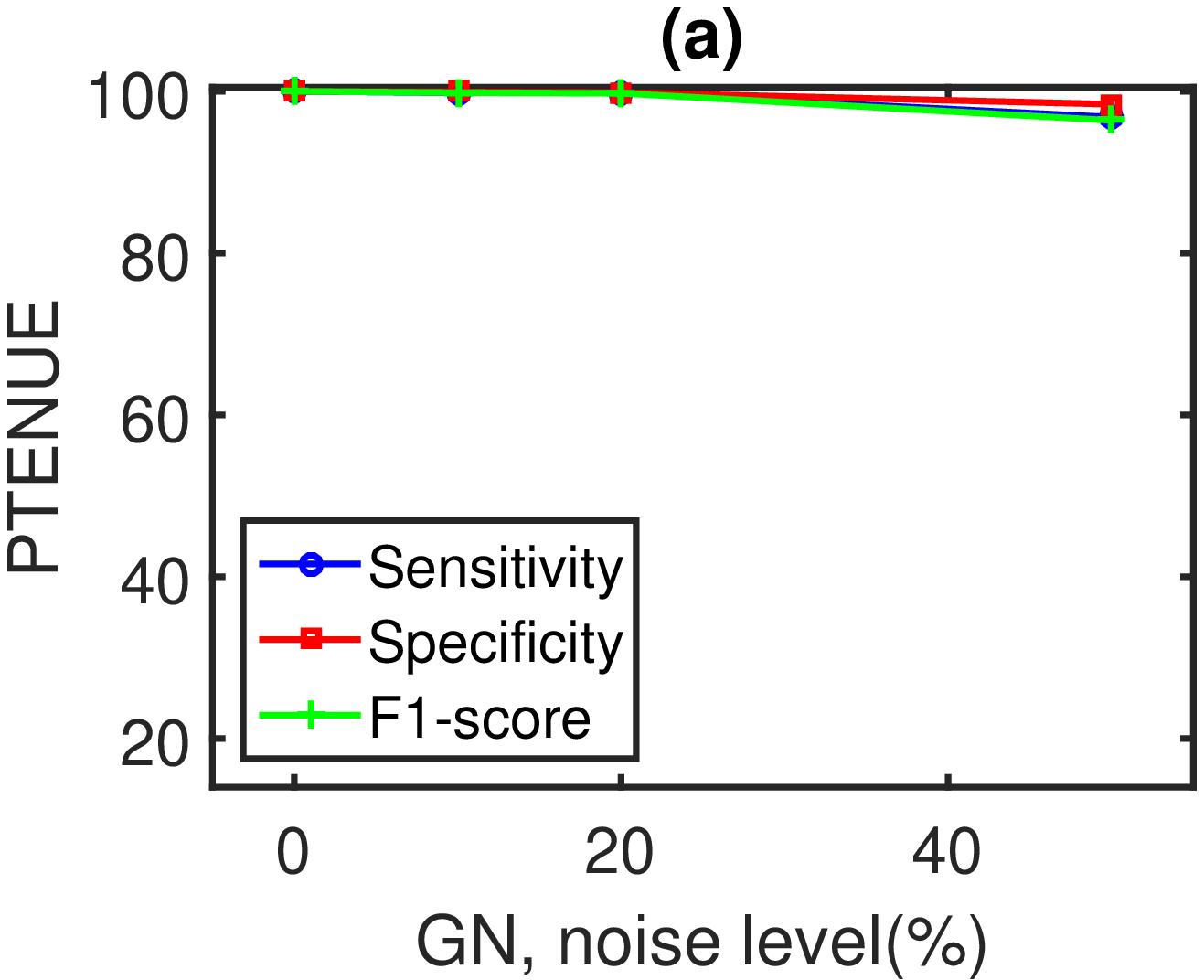}          
\includegraphics[scale=.26]{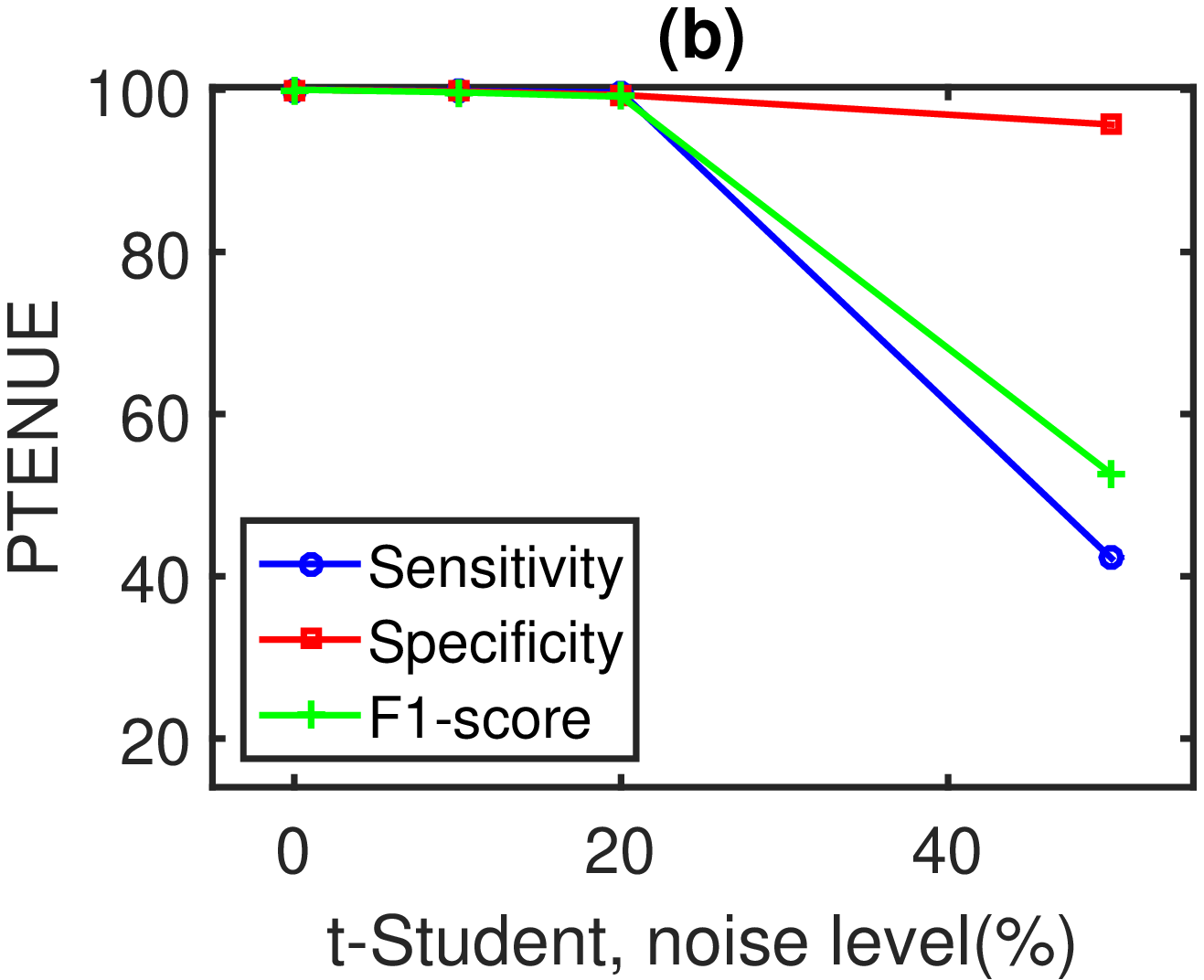}         
\includegraphics[scale=.26]{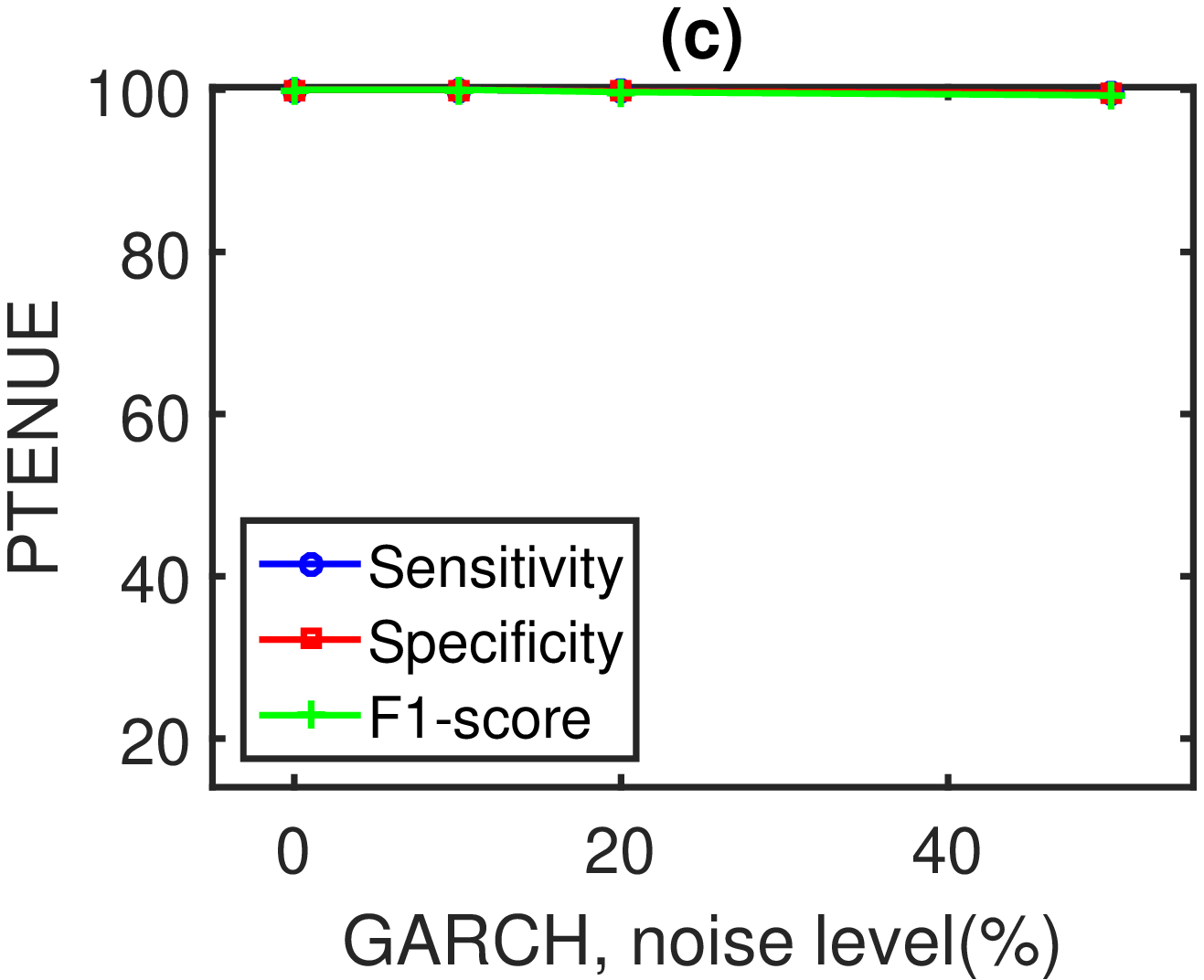}   
\includegraphics[scale=.26]{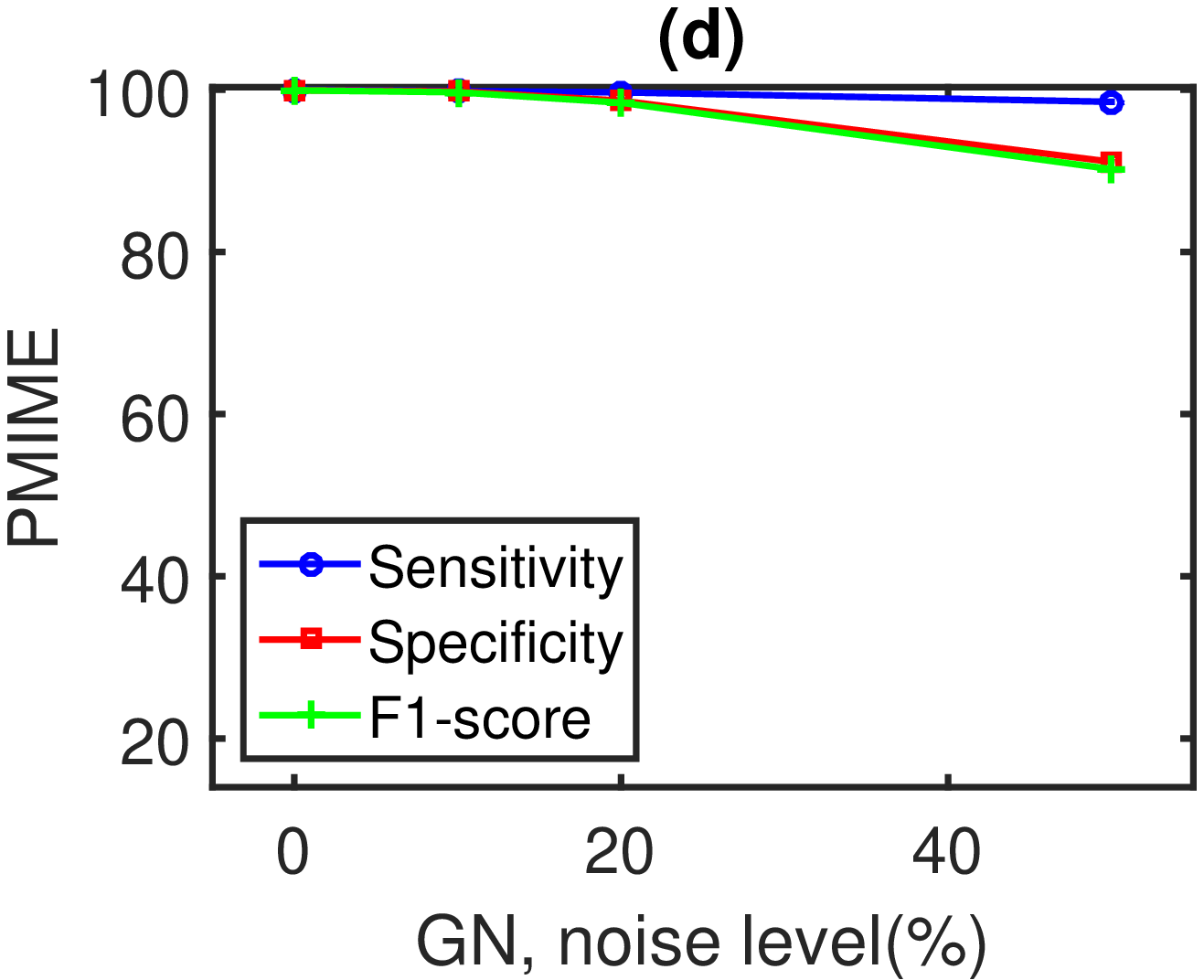}            
\includegraphics[scale=.26]{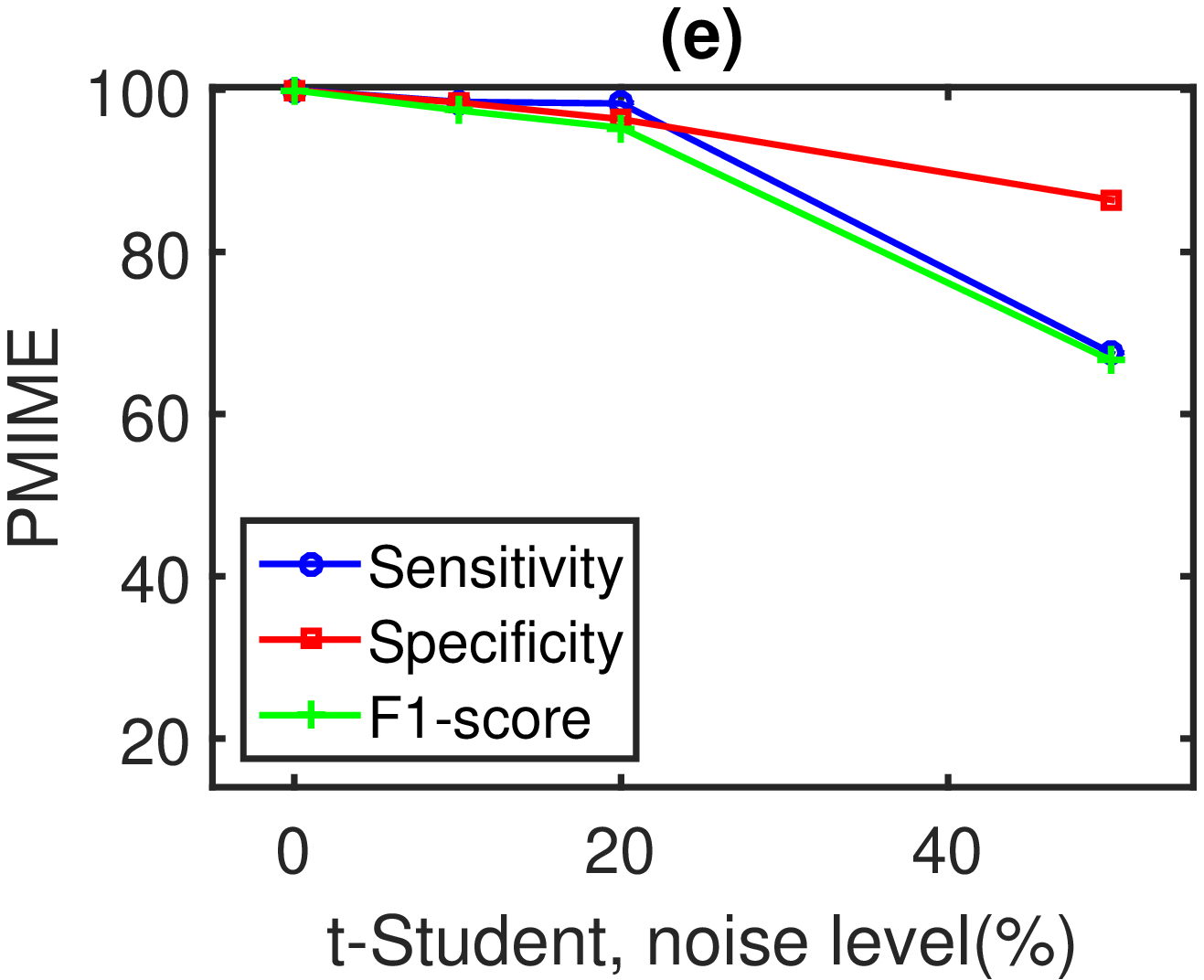}           
\includegraphics[scale=.26]{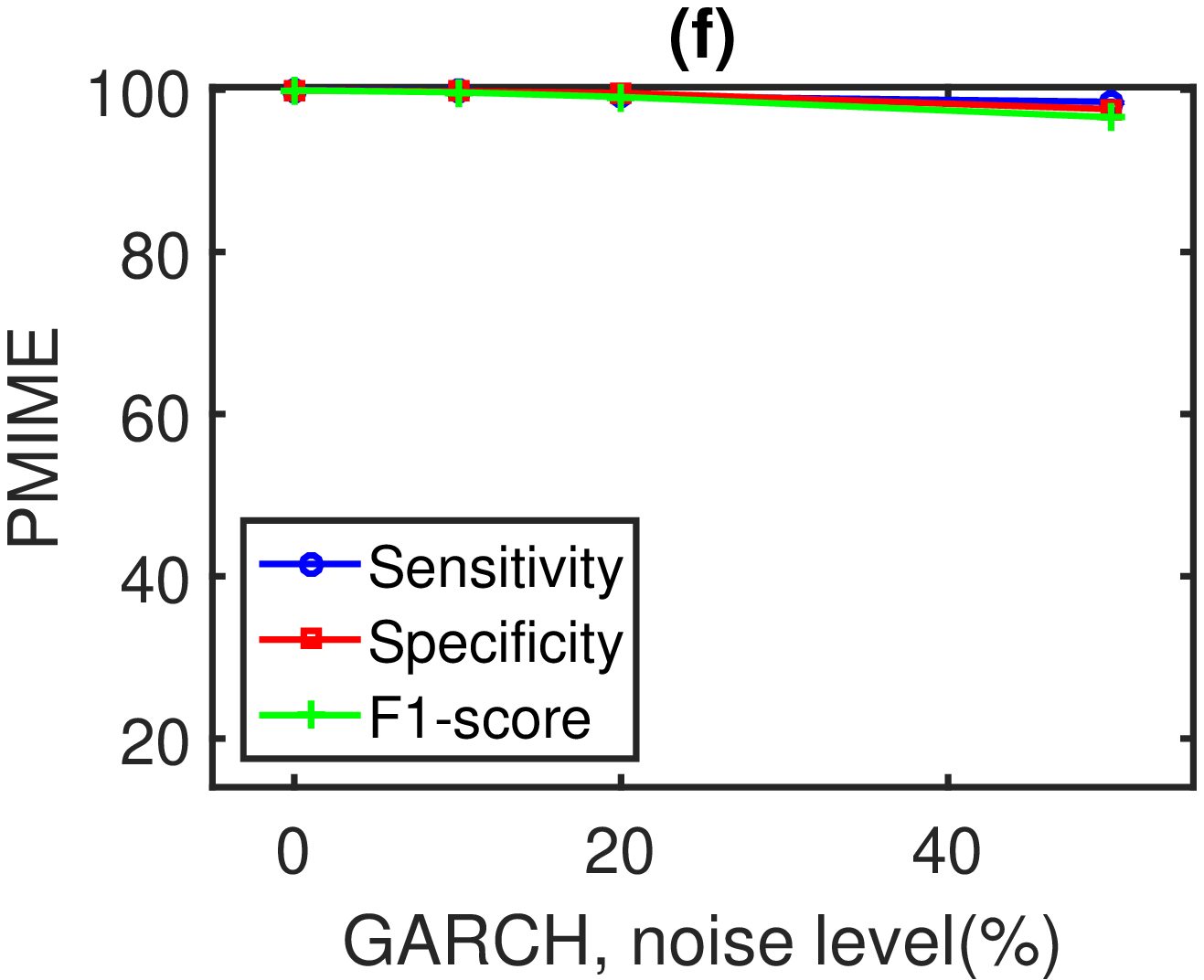}     
\includegraphics[scale=.26]{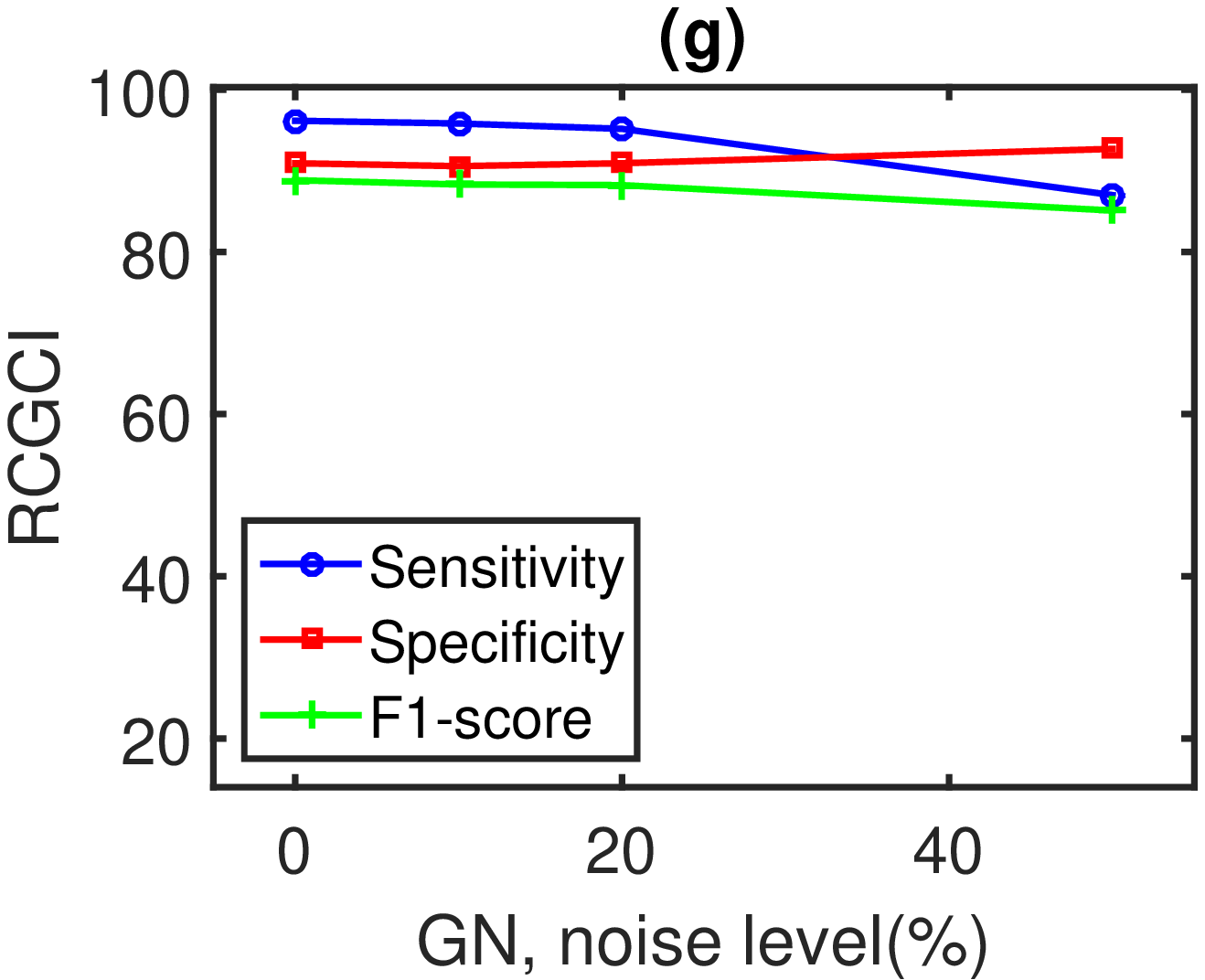}            
\includegraphics[scale=.26]{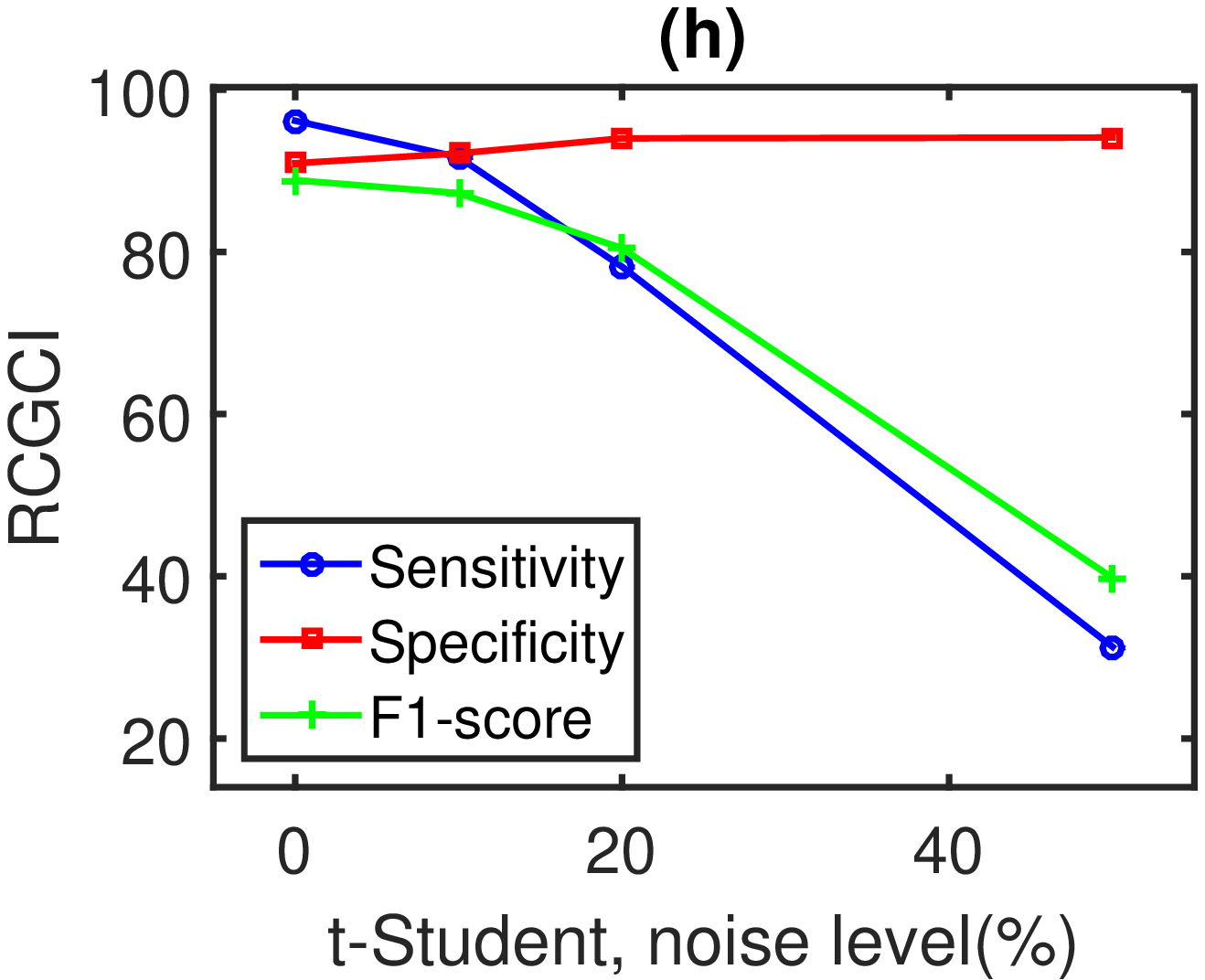}           
\includegraphics[scale=.26]{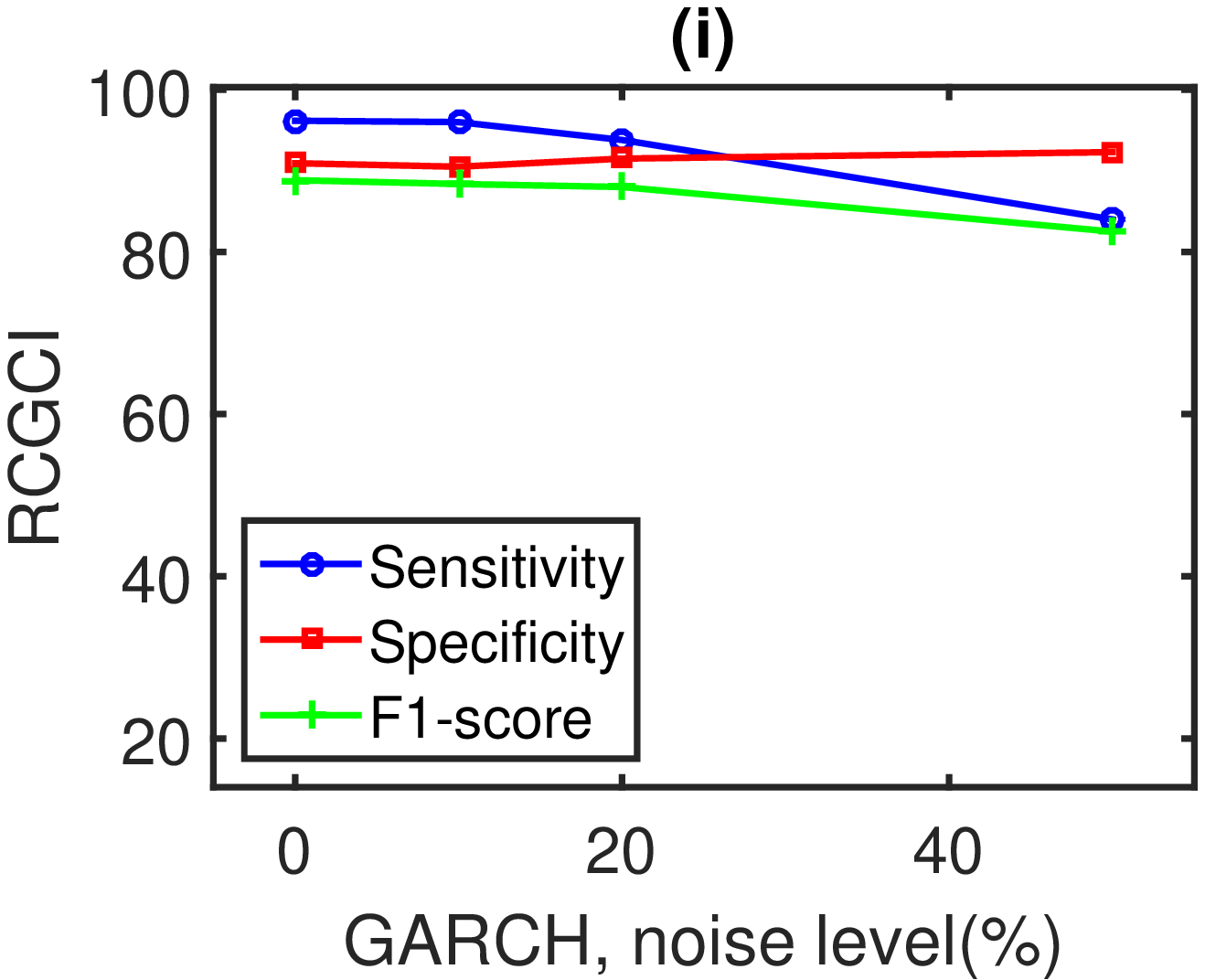}      
\includegraphics[scale=.26]{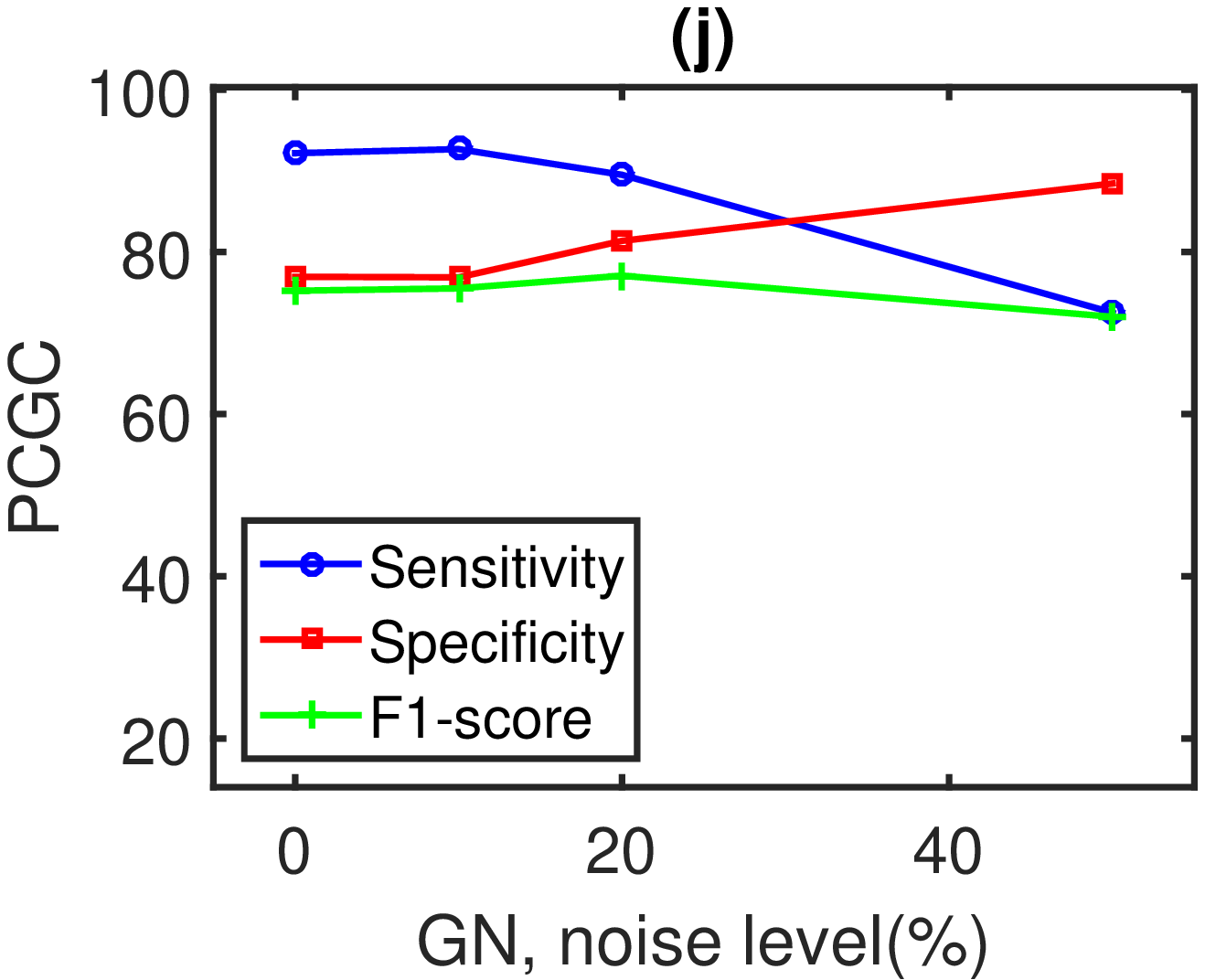}             
\includegraphics[scale=.26]{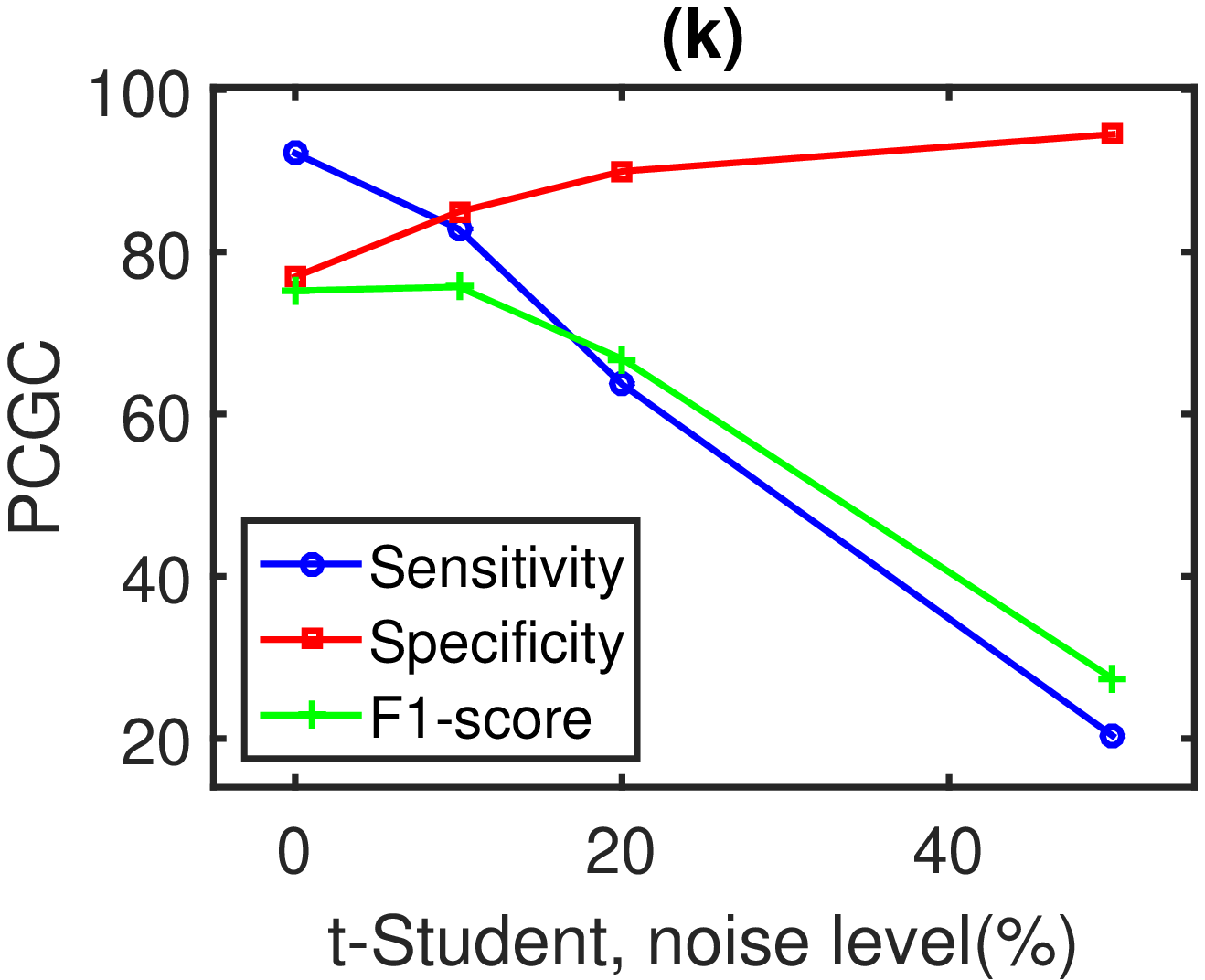}             
\includegraphics[scale=.26]{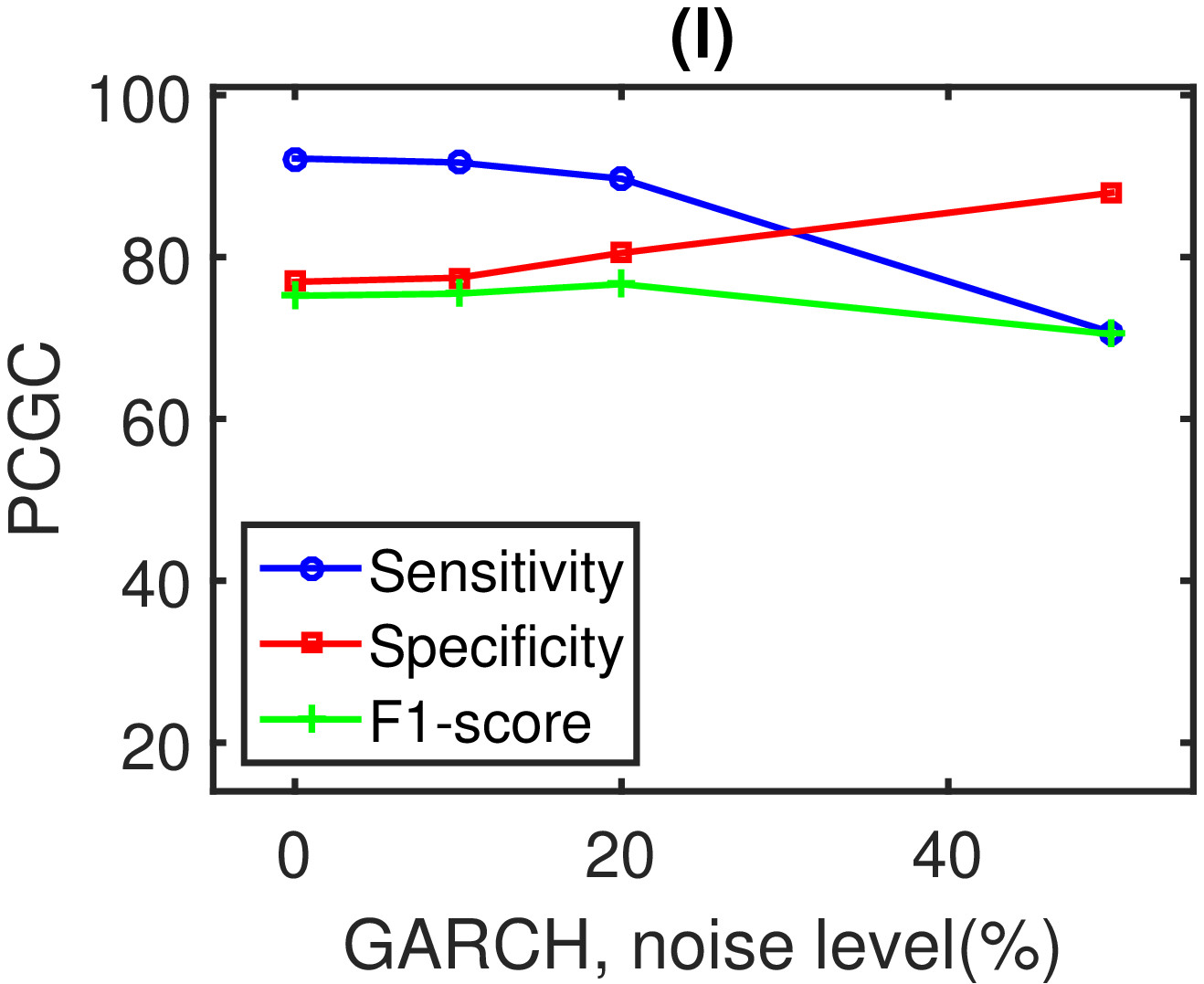}       
 \caption{Accuracy indexes for S3 in $K=5$ variables and $c=0.2$, with increasing amplitude levels
 of observational noise for the three types of noise: Gaussian white noise (left column), t-Student (middle column) and GARCH(1,1) (right column), and for the four causality measures: PTENUE (first row), PMIME (second row), RCGCI (third row) and PCGC (fourth row).}
 \label{fig:S3noiselev}
\end{figure}

Figure~\ref{fig:S3noiselev} shows the results of the effect of the three noise types on the four causality measures for increasing noise levels (we set $n=2048$).
The nonlinear measures are more stable to the effect of noise than the linear measures. For small noise levels, the
performance of the measures is rather stable, but as the noise level rises, sensitivity decreases and therefore their
overall performance gets worse. On the other hand, specificity slightly increases as the noise level grows showing the tendency of the measures to detect less links with the increase of noise level. The
addition of noise from t-Student distribution seems to significantly affect all the measures, especially when noise level is high, where the PMIME is the least affected. Interestingly, the GARCH noise seems to affect only the linear measures, which shows high robustness to volatility clustering of the nonlinear measures.

\section{Conclusions}
\label{sec:Conclusions}

In this study, a comparative analysis has been performed in order to evaluate an ensemble of well known causality
measures. The robustness of the different measures is investigated on artificial data from multivariate systems with
different characteristics. The main contribution of this study is the evaluation of an ensemble of different causality
measures that have not been previously compared on simulation systems with different characteristics.
Further, the study focuses on truly high dimensional systems of up to 100 variables.
Finally, the noise robustness is examined, which is an essential part of this work since many 
comparative studies do not include noisy samples.

The evaluation of the causality measures showed differences in their performance in the different
systems. First of all, the empirical results confirmed some generally recognized facts:
a) the linear causality measures cannot systematically infer about the
nonlinear causal effects and therefore their overall performance is worse than that of the nonlinear ones,
b) bivariate causality measures indicate both direct and indirect causal effects while false indications of directed relationships often arise,
c) fully conditioned causality measures are the most affected ones by the curse of dimensionality.

The main new findings of this study can be summarized to the following points:
\begin{itemize}

\item Partially conditioned measures (applying dimension reduction) are superior to bivariate and fully conditioned causality measures in both low and high dimensional systems. The utilization of the dimension reduction techniques to address the curse of the dimensionality seems indispensable when direct causality in multivariate time series is to be estimated. 
 
\item In total, PTENUE and PMIME are the optimal causality measures with similar performance at most cases.
They seem to be robust, unaffected by the curse of dimensionality, the nature and the strength of the connections
and the type of noise. Sufficiently long time series are though required to achieve their optimal performance.
PTENUE is more robust in case of the stochastic systems; PMIME estimates slightly higher percentages than the
nominal level ($5\%$) for the uncoupled pairs of variables and therefore its specificity is lower compared to PTENUE's.

\item The nonlinear measures are more stable to the effect of noise than the linear measures.
In particular, the GARCH noise seems to affect only the linear measures.
PMIME seems to be more robust to high levels of noise, especially of heavy tailed distribution.

\end{itemize}

It remains to be studied whether the optimal causality measures perform well also in systems with denser 
causality structures.
Further, as a future work we aim to examine the performance of PTENUE and PMIME in comparison to different causal
discovery methods, such as the Fast Causal Inference algorithm \cite{Entner10}. The effect of latent variables and 
instantaneous causality on the performance of the causality measures are also factors to be examined in future work.

\section*{Acknowledgments}

This project received funding from the Hellenic Foundation for Research and Innovation (HFRI) and
the General Secretariat for Research and Technology (GSRT), under Grant Agreement No. 794.

\bibliographystyle{unsrt}
\bibliography{bib_DirectCausality}

\end{document}